\documentclass[10pt,twocolumn]{article}

\usepackage[margin=1in]{geometry}
\usepackage{graphicx}
\usepackage{booktabs}
\usepackage{amsmath,amssymb}
\usepackage[ruled,vlined,linesnumbered]{algorithm2e}
\usepackage[colorlinks=true,linkcolor=blue,citecolor=blue,urlcolor=blue]{hyperref}
\usepackage{xcolor}
\usepackage{multirow}
\usepackage{subcaption}
\usepackage[round]{natbib}
\usepackage{listings}
\usepackage{longtable}
\usepackage{float}
\usepackage{enumitem}
\usepackage{tabularx}

\title{UniRank: A Multi-Agent Calibration Pipeline for Estimating\\University Rankings from Anonymized Bibliometric Signals}
\author{
  Pedram Riyazimehr\textsuperscript{*} \quad Seyyed Ehsan Mahmoudi\\[4pt]
  NotionWave\\
  \texttt{pedram.riazi@notionwave.com}
}
\date{February 2026}

\begin{document}
\maketitle

\begin{abstract}
We present UniRank, a multi-agent LLM pipeline that estimates university positions across global ranking systems using only publicly available bibliometric data from OpenAlex and Semantic Scholar. The system employs a three-stage architecture: (a)~zero-shot estimation from \emph{anonymized} institutional metrics, (b)~per-system tool-augmented calibration against real ranked universities, and (c)~final synthesis. Critically, institutions are anonymized---names, countries, DOIs, paper titles, and collaboration countries are all redacted---and their actual ranks are hidden from the calibration tools during evaluation, preventing LLM memorization from confounding results.

On the Times Higher Education (THE) World University Rankings ($n{=}352$), the system achieves MAE${}=251.5$ rank positions, Median~AE${}=131.5$, PNMAE${}=12.03\%$, Spearman~$\rho{=}0.769$, Kendall~$\tau{=}0.591$, hit rate @50${}=20.7\%$, hit rate @100${}=39.8\%$, and a Memorization Index of exactly zero (no exact-match zero-width predictions among all 352 universities). The systematic positive signed error ($+190.1$ positions, indicating the system consistently predicts worse ranks than actual) and monotonic performance degradation from elite tier (MAE${}=60.5$, hit@100${}=90.5\%$) to tail tier (MAE${}=328.2$, hit@100${}=20.8\%$) provide strong evidence that the pipeline performs genuine analytical reasoning rather than recalling memorized rankings. A live demo is available at \url{https://unirank.scinito.ai}.
\end{abstract}

\textbf{Keywords:} university rankings, multi-agent systems, LLM evaluation, bibliometrics, anonymization, tool-augmented reasoning, OpenAlex, ranking estimation, decontamination, MAgICoRe

\section{Introduction}\label{sec:intro}

\subsection{Problem Statement}

Global university rankings---the QS World University Rankings, Times Higher Education (THE) World University Rankings, and the Academic Ranking of World Universities (ARWU)---are among the most influential instruments in higher education \citep{hazelkorn2015rankings}. They shape student enrollment decisions, institutional funding allocations, government policy, and cross-border academic partnerships \citep{marginson2014rankings}. Yet these rankings rely heavily on data that is expensive to collect, partially proprietary, and methodologically opaque:

\begin{itemize}[nosep]
  \item \textbf{Survey data}: QS allocates 45\% of its weight to academic and employer reputation surveys \citep{qs2026}; THE allocates ${\sim}18\%$ to research reputation \citep{the2026}.
  \item \textbf{Institution-reported data}: Student--staff ratios, international student counts, and financial data are self-reported with limited independent verification.
  \item \textbf{Proprietary scoring}: Each system uses different weights and normalization procedures, producing substantially different rankings for the same institution.
\end{itemize}

These rankings cover only ${\sim}1{,}500$ of the world's ${\sim}30{,}000$+ higher education institutions, leaving the vast majority unranked \citep{marginson2014rankings}.

\subsection{Research Question}

Can a multi-agent LLM pipeline estimate university ranking positions from \textbf{publicly available bibliometric data alone}, without access to survey data, reputation signals, or proprietary institutional data?

\subsection{Why This Is Not Simple LLM Memorization}

A na\"ive approach---``Ask GPT where MIT ranks''---fails for three reasons: (1)~LLMs have seen ranking lists in their training data, making any direct query a memorization exercise \citep{carlini2021extracting}; (2)~training data is temporally stale; (3)~a memorized rank provides no methodological insight. Our approach addresses all three:

\begin{itemize}[nosep]
  \item \textbf{Anonymization}: The LLM receives only an opaque identifier (e.g., \texttt{INST-A3F2B1C4}) with numeric metrics. All identifying information---institution name, country, paper titles, DOIs, collaboration countries---is redacted.
  \item \textbf{Data hiding}: During evaluation, the target university is physically removed from the ranking data store, so calibration tools cannot return the ground-truth answer.
  \item \textbf{Tool-augmented reasoning}: The LLM must use tools to compare the anonymous institution's metrics against real ranked universities---it cannot retrieve a memorized ranking.
\end{itemize}

\subsection{Contributions}

\begin{enumerate}[nosep]
  \item A novel three-stage multi-agent architecture (MAgICoRe-inspired) for ranking estimation.
  \item A comprehensive anonymization and data-hiding protocol that isolates LLM reasoning from memorized knowledge.
  \item A bibliometric feature set (16 indicators) derived entirely from open data sources (OpenAlex, Semantic Scholar).
  \item An evaluation framework with stratified test sets, multiple accuracy metrics, and Wilson confidence interval-based accuracy claims.
  \item The \textbf{Memorization Index (MI)}---a novel metric for quantifying evidence of LLM memorization in estimation tasks.
  \item A live, publicly accessible demo at \url{https://unirank.scinito.ai}.
\end{enumerate}

\subsection{Paper Organization}

Section~\ref{sec:related} reviews related work. Section~\ref{sec:arch} describes the system architecture. Section~\ref{sec:eval} presents the evaluation framework. Section~\ref{sec:results} reports results on the THE ranking system. Section~\ref{sec:discussion} discusses memorization evidence and error analysis. Section~\ref{sec:future} outlines future work, and Section~\ref{sec:conclusion} concludes.

\section{Related Work}\label{sec:related}

\subsection{University Ranking Systems}

The three major global ranking systems differ substantially in methodology. QS allocates 50\% to research metrics, 45\% to reputation surveys, and 5\% to sustainability \citep{qs2026}. THE emphasizes research quality (30\%), research environment (29.5\%), teaching (15\%), international outlook (7.5\%), and industry impact (4\%) \citep{the2026}. ARWU uses exclusively hard research metrics: highly cited researchers (20\%), Nature/Science publications (20\%), total publications (20\%), Nobel/Fields winners (30\%), and per-capita performance (10\%) \citep{arwu2025}. The divergence between systems means the same university can rank very differently depending on how heavily reputation, teaching, and research are weighted \citep{hazelkorn2015rankings, marginson2014rankings}.

\subsection{Bibliometric Analysis and Scientometrics}

OpenAlex provides an open catalog of 240M+ scholarly works, 100K+ institutions, and their bibliometric relationships \citep{priem2022openalex}. Semantic Scholar contributes the influential citation metric---identifying citations that meaningfully impact the citing paper's methodology or results \citep{valenzuela2015identifying}. The h-index \citep{hirsch2005index} remains widely used despite known limitations (field dependence, career-stage bias). Field-Weighted Citation Impact (FWCI) provides field-normalized impact measurement, and we use the closely related 2-year mean citedness from OpenAlex as a proxy.

\subsection{LLMs for Knowledge-Intensive Tasks}

Tool-augmented LLMs such as ReAct \citep{yao2023react} and Toolformer \citep{schick2023toolformer} demonstrate that LLMs can perform complex reasoning when equipped with external tools. Multi-agent architectures---AutoGen \citep{wu2024autogen}, CAMEL \citep{li2023camel}, MetaGPT \citep{hong2024metagpt}---show that distributing tasks across specialized agents improves performance on complex problems. MAgICoRe (Multi-Agent, Iterative, Coarse-to-Fine Refinement) \citep{chen2025magicore} provides the direct architectural inspiration for our pipeline, demonstrating that iterative refinement across agents produces better results than single-pass estimation.

\subsection{LLM Evaluation and Decontamination}

The risk of training data contamination in LLM evaluation is well-documented \citep{carlini2021extracting, magar2022data}. Our anonymization-based approach complements existing decontamination techniques (n-gram filtering, canary insertion) by operating at the input level: the LLM never sees identifiable information, making memorization-based shortcuts impossible regardless of what the model has seen during training.

\section{System Architecture}\label{sec:arch}

\subsection{Architecture Overview}

\begin{figure*}[t]
\centering
\includegraphics[width=\textwidth]{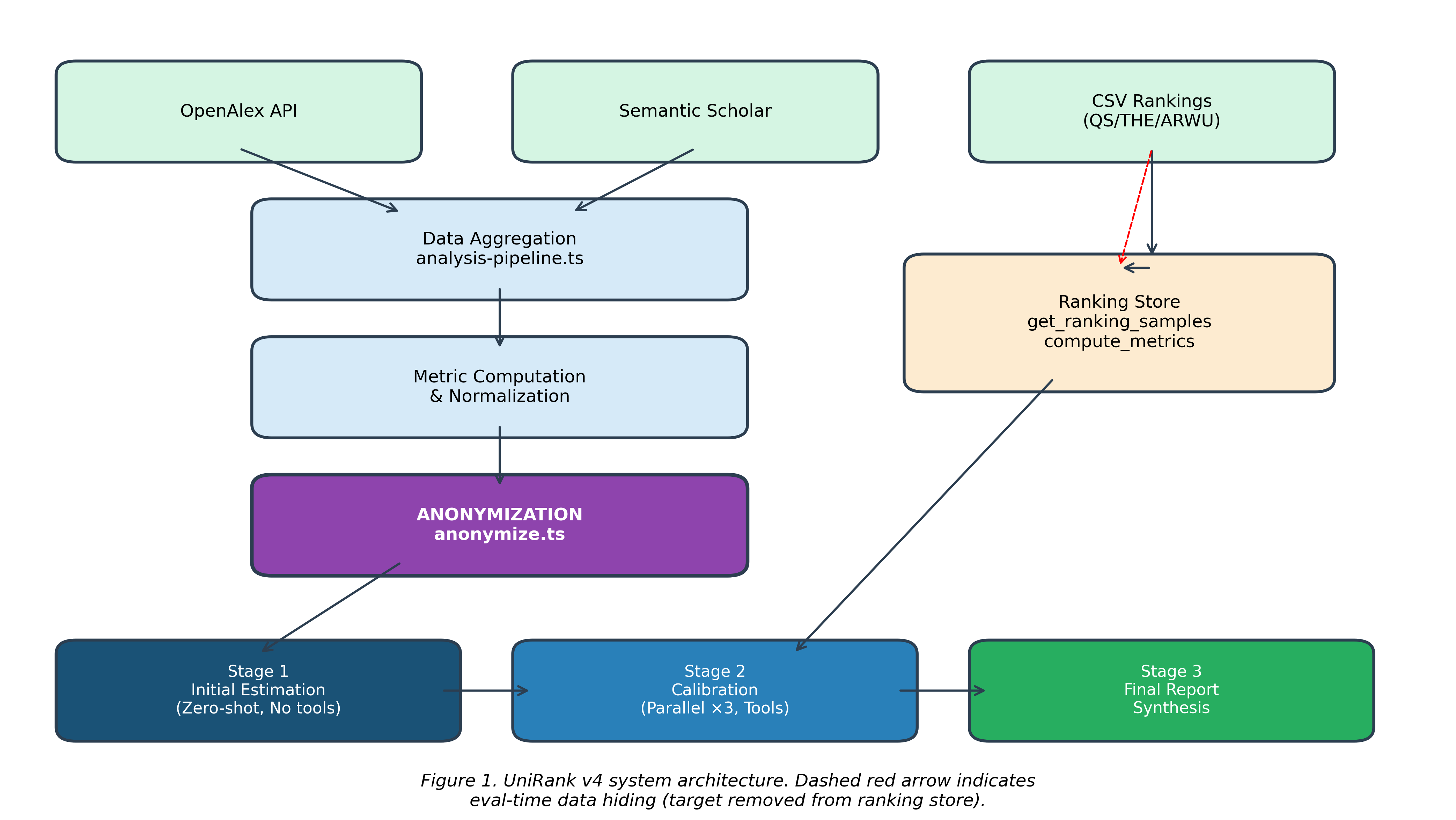}
  \caption{UniRank system architecture. Data from OpenAlex and Semantic Scholar is aggregated, normalized, and anonymized before entering the three-stage LLM pipeline. During evaluation, the target university is hidden from the ranking store (dashed line) to prevent data leakage.}
  \label{fig:architecture}
\end{figure*}

Figure~\ref{fig:architecture} presents the end-to-end UniRank pipeline. Data from OpenAlex and Semantic Scholar is aggregated, normalized, and anonymized before entering the three-stage LLM pipeline. During evaluation, the target university is hidden from the ranking store to prevent data leakage.

\subsubsection{Formal Problem Definition}

Let $\mathcal{U}$ be the set of all universities. For each ranking system $s \in \{QS, THE, ARWU\}$, let $\pi_s : \mathcal{U} \to \mathbb{N}$ be the ground-truth ranking function. For a target university $u^* \in \mathcal{U}$, we observe a feature vector $\mathbf{x}_{u^*} \in \mathbb{R}^d$ computed from publicly available bibliometric data, where $d = 16$ features (Table~\ref{tab:features}).

For each system $s$, we produce a predicted rank range $[\hat{r}_s^{\min}, \hat{r}_s^{\max}]$ and a point estimate $\hat{r}_s = (\hat{r}_s^{\min} + \hat{r}_s^{\max}) / 2$. Let $\alpha : \mathcal{U} \to \mathcal{U}'$ be the anonymization function that maps $u^*$ to an opaque identifier while preserving only $\mathbf{x}_{u^*}$. During evaluation of $u^*$, the ranking store $\mathcal{R}_s$ is modified to $\mathcal{R}_s \setminus \{u^*\}$.

\textbf{Objective}: Minimize $\text{MAE} = \frac{1}{|\mathcal{T}|}\sum_{u \in \mathcal{T}} |\hat{r}_s(u) - \pi_s(u)|$ over a stratified test set $\mathcal{T} \subset \mathcal{U}$, subject to the anonymization and data-hiding constraints.

\subsection{Data Sources}

\textbf{OpenAlex} (primary): We query institution profiles, disciplinary distributions, international collaboration counts, open access breakdowns, top cited works with FWCI, and research excellence counts (top 10\% by FWCI). All queries are filtered to publication years 2020--2025 \citep{priem2022openalex}.

\textbf{Semantic Scholar} (enrichment): For the top 10 cited works, we retrieve influential citation counts and ratios via the batch endpoint \citep{valenzuela2015identifying}. The influential citation ratio (influential/total) provides a quality signal unavailable from OpenAlex.

\textbf{Ground truth rankings}: QS 2026, THE 2024, and ARWU 2025 data from publicly available CSV datasets serve as both calibration anchors and evaluation ground truth.

\subsection{Metric Computation and Normalization}

Table~\ref{tab:features} lists the 16 bibliometric indicators computed for each institution.

\begin{table}[t]
\centering
\caption{Bibliometric feature set (16 indicators).}
\label{tab:features}
\small
\begin{tabular}{@{}lll@{}}
\toprule
\textbf{Metric} & \textbf{Source} & \textbf{Description} \\
\midrule
worksCount & OA & Total publications \\
citedByCount & OA & Total citations \\
hIndex & OA & Hirsch index \\
i10Index & OA & Papers with 10+ citations \\
2yr mean citedness & OA & FWCI proxy \\
citationsPerWork & OA & Average citations/paper \\
5yr works growth & OA & Publication growth (\%) \\
5yr citation growth & OA & Citation growth (\%) \\
researchExcellence\% & OA & \% in global top 10\% FWCI \\
intlCollaboration\% & OA & \% with intl.\ co-authors \\
openAccess\% & OA & Open access rate \\
disciplinaryBreadth & OA & Shannon entropy \\
normalizedResearch & OA & 0--100 research score \\
normalizedImpact & OA & 0--100 impact score \\
normalizedExcellence & OA & 0--100 excellence score \\
influentialRatio & S2 & Influential/total citations \\
\bottomrule
\end{tabular}
\end{table}

Each metric is normalized against global tier benchmarks using:
\begin{equation}\label{eq:norm}
\text{score} = \text{clamp}\!\left(\frac{v - \text{floor}}{\text{ceiling} - \text{floor}} \times 100,\; 0,\; 100\right)
\end{equation}
where $\text{floor} = p_{25}$ of Tier~4 (Top~500) and $\text{ceiling} = p_{75}$ of Tier~1 (Top~10). For each metric, we additionally compute a Z-score relative to the closest tier:
\begin{equation}\label{eq:zscore}
z = \frac{v - \mu_{\text{closest}}}{\sigma_{\text{closest}}}
\end{equation}
producing signals like ``h-index: $+1.2\sigma$ from Tier~2 mean.''

\subsection{Three-Stage Multi-Agent Pipeline}

\begin{figure*}[t]
  \centering
  \includegraphics[width=\textwidth]{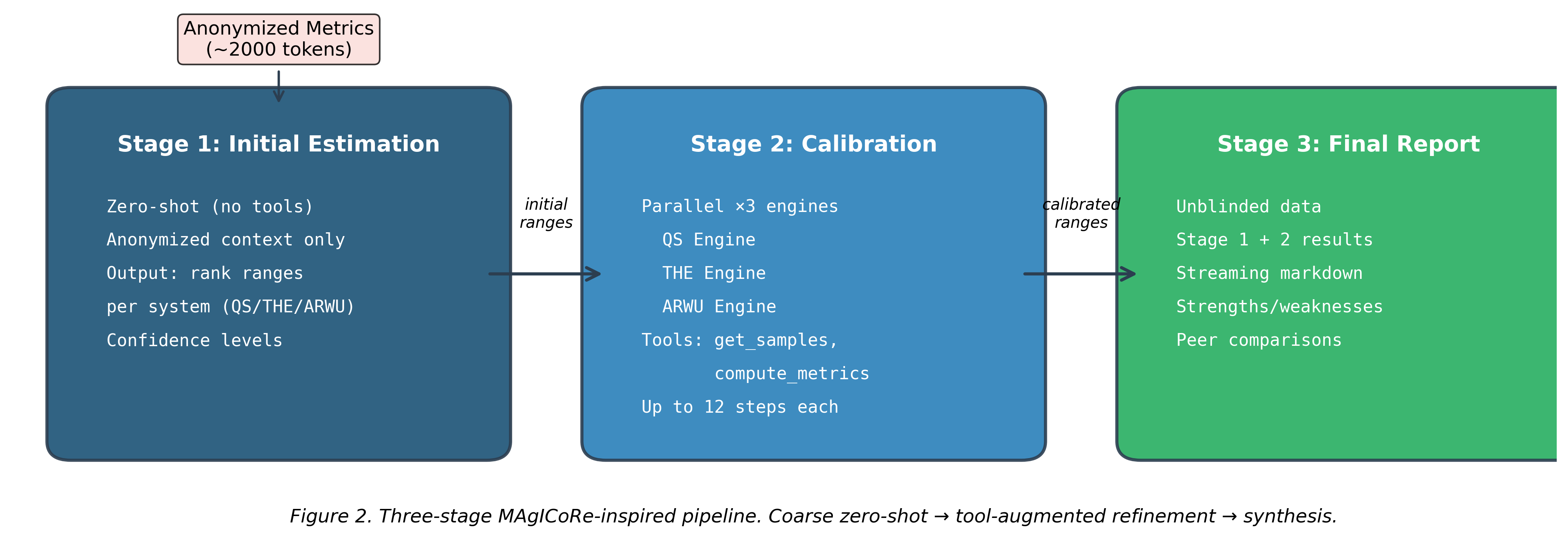}
  \caption{Three-stage pipeline: Stage~1 produces coarse zero-shot estimates from anonymized metrics. Stage~2 refines per-system with tool-augmented calibration (parallel). Stage~3 synthesizes the final report.}
  \label{fig:pipeline}
\end{figure*}

Figure~\ref{fig:pipeline} illustrates the pipeline flow, inspired by MAgICoRe's coarse-to-fine refinement pattern \citep{chen2025magicore}.

\textbf{Stage~1: Initial Zero-Shot Estimation.} The LLM receives anonymized metrics (${\sim}2{,}000$ tokens) and produces rank ranges for each system. No tools are available; estimation relies solely on the model's understanding of each ranking methodology and the provided Z-scores.

\textbf{Stage~2: Per-System Calibration (Parallel).} Three independent calibration engines run in parallel (one per system). Each has access to two tools:
\begin{itemize}[nosep]
  \item \texttt{get\_ranking\_samples(system, rankMin, rankMax, count)}: Returns real universities within the specified rank range, including names, actual ranks, and official sub-scores. During evaluation, the target is hidden from this data.
  \item \texttt{compute\_metrics(universityName)}: Computes the full bibliometric feature set for a named university using the same pipeline, enabling direct metric comparison.
\end{itemize}
Each engine executes up to 12 agentic steps: fetching samples, computing metrics, comparing, adjusting, and producing a calibrated range (target width: 30--50 positions).

\textbf{Stage~3: Final Report Synthesis.} This stage receives the full (non-anonymized) university data along with Stage~1 and Stage~2 results, producing a structured analysis report.

\subsection{Anonymization Protocol}\label{sec:anon}

This is the core contribution addressing memorization concerns. The anonymization function transforms all identifying information before any LLM interaction (Figure~\ref{fig:anonymization}):

\begin{figure*}[t]
  \centering
  \includegraphics[width=\textwidth]{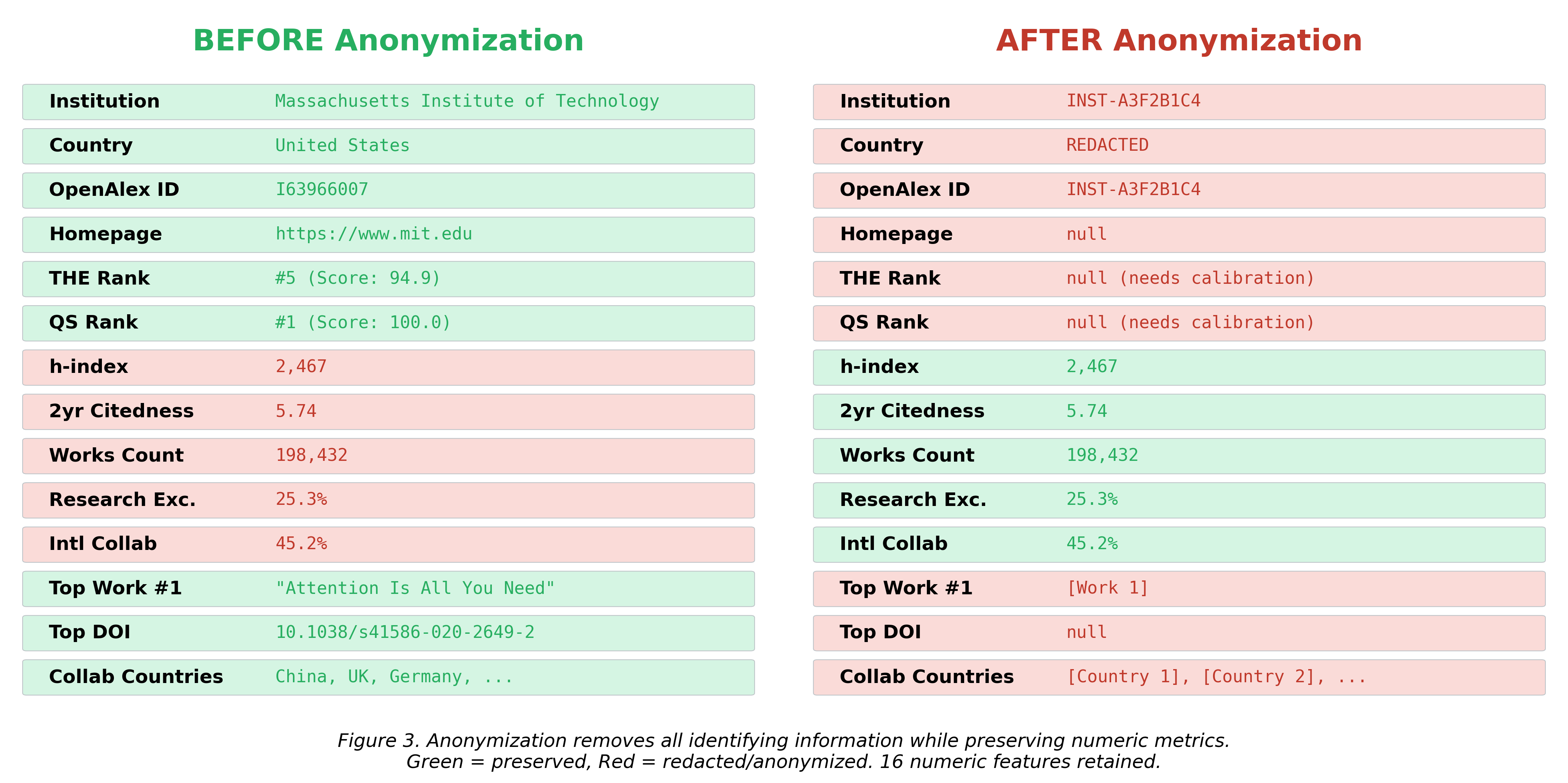}
  \caption{Anonymization before/after comparison. All identifying information (name, country, DOIs, paper titles, collaboration countries) is redacted; only numeric metrics are preserved.}
  \label{fig:anonymization}
\end{figure*}

\begin{itemize}[nosep]
  \item Institution name $\to$ random hex identifier (\texttt{INST-A3F2B1C4})
  \item Country $\to$ \texttt{REDACTED}
  \item Paper titles $\to$ \texttt{[Work 1]}, \texttt{[Work 2]}, \ldots
  \item DOIs $\to$ \texttt{null}
  \item Collaboration countries $\to$ \texttt{[Country 1]}, \texttt{[Country 2]}, \ldots
  \item Published rankings (QS/THE/ARWU) $\to$ \texttt{null}
  \item Semantic Scholar TLDRs $\to$ \texttt{null}
\end{itemize}

\textbf{Preserved fields} (numeric only): all computed metrics, Z-scores, tier-relative positioning, yearly trends, field distribution percentages, citation counts, FWCI, citation percentiles, and influential citation data.

\subsection{Model Configuration}

The pipeline uses GPT-5.2 (via \texttt{@ai-sdk/openai}) with reasoning effort set to \texttt{medium}. Structured outputs are enforced via Zod schemas through the Vercel AI SDK. The Helicone AI Gateway provides logging and monitoring.

\section{Evaluation Framework}\label{sec:eval}

\subsection{Test Set Construction}

The test set comprises 500 universities with OpenAlex ID resolution, constructed via stratified sampling across two dimensions:

\textbf{Tier stratification}: Elite (ranks 1--25, 20\%), Strong (26--150, 20\%), Mid (151--400, 20\%), Lower (401--700, 20\%), Tail (700+, 20\%). Tier assignment is based on consensus rank (median of available QS/THE/ARWU ranks).

\textbf{Regional stratification}: North America (${\sim}20\%$), Europe (${\sim}25\%$), Asia-Pacific (${\sim}25\%$), Rest of World (${\sim}15\%$), Mixed/Other (${\sim}15\%$).

\subsection{Double-Blind Evaluation Protocol}

For each target university, evaluation proceeds in four steps: (1)~\emph{Data hiding}: the target's record is physically removed from the in-memory ranking store, verified before proceeding; (2)~\emph{Anonymization}: the standard protocol (Section~\ref{sec:anon}) strips all identifying information; (3)~\emph{Pipeline execution}: the three-stage pipeline runs against the anonymized, hidden data; (4)~\emph{Restoration}: the hidden university is restored to the ranking store.

\subsection{Evaluation Metrics}

\subsubsection{Primary Metrics}

\begin{align}
\text{MAE} &= \frac{1}{n}\sum_{i=1}^{n} |m_i - a_i| \label{eq:mae} \\
\text{Median AE} &= \text{median}(|m_i - a_i|) \\
\text{PNMAE} &= \frac{1}{n}\sum_{i=1}^{n} \left|\frac{m_i - 1}{N-1} - \frac{a_i - 1}{N-1}\right| \times 100 \label{eq:pnmae}
\end{align}
where $m_i$ is the predicted midpoint, $a_i$ is the actual rank, and $N$ is the total number of ranked universities in the system.

\subsubsection{Hit Rate Metrics}

\begin{equation}
\text{Hit@}k = \frac{1}{n}\sum_{i=1}^{n} \mathbb{1}[|m_i - a_i| \leq k] \times 100
\end{equation}
for $k \in \{25, 50, 100\}$.

\subsubsection{Range Metrics}

Range coverage measures the fraction of actual ranks falling within the predicted range: $\frac{1}{n}\sum \mathbb{1}[r_{\min} \leq a_i \leq r_{\max}]$. Mean range width is $\frac{1}{n}\sum (r_{\max} - r_{\min})$.

\subsubsection{Memorization Index (MI)}

To explicitly quantify evidence of memorization:
\begin{equation}\label{eq:mi}
MI = \frac{|\{u \in \mathcal{T} : AE(u) = 0 \wedge W(u) = 0\}|}{|\mathcal{T}|}
\end{equation}
where $W(u) = \hat{r}^{\max}(u) - \hat{r}^{\min}(u)$ is the range width. A non-zero MI indicates suspiciously perfect predictions with zero-width ranges---a hallmark of memorized retrieval. We expect MI~$\approx 0$ for a reasoning-based system.

\subsubsection{Correlation and Error Decomposition}

Spearman's $\rho$, Pearson's $r$, and Kendall's $\tau$ measure ordinal and linear correlation. We additionally report RMSE, signed error (directional bias), calibration slope $\beta$ from OLS regression ($\hat{r} = \alpha + \beta \cdot r$; $\beta = 1$ indicates perfect calibration), and Cohen's $\kappa$ for tier classification agreement.

\subsection{Accuracy Claims with Confidence Intervals}

To avoid overclaiming, we compute Wilson score interval lower bounds \citep{wilson1927probable}:
\begin{equation}\label{eq:wilson}
L = \frac{\hat{p} + \frac{z^2}{2n} - z\sqrt{\frac{\hat{p}(1-\hat{p}) + \frac{z^2}{4n}}{n}}}{1 + \frac{z^2}{n}}
\end{equation}
with $z = 1.96$ (95\% CI). The claimed accuracy is $\lfloor L \times 100 \rfloor\%$.

\subsection{Statistical Testing Plan}

\begin{itemize}[nosep]
  \item \textbf{Wilcoxon signed-rank}: Compare initial vs.\ calibrated absolute errors (paired).
  \item \textbf{McNemar's test}: Compare hit rates before/after calibration.
  \item \textbf{Kruskal--Wallis}: Compare errors across tiers.
\end{itemize}
All reported metrics include 95\% bootstrapped confidence intervals ($B{=}10{,}000$ resamples).

\section{Results}\label{sec:results}

We present results on the THE World University Rankings---our primary evaluation target. The evaluation ran on 357 universities from the THE stratified test set; 352 produced successful predictions and 5 failed (pipeline errors, 1.4\% failure rate).

\subsection{Aggregate Metrics}

Table~\ref{tab:aggregate} reports aggregate performance over $n{=}352$ successful predictions. THE contains approximately 2{,}092 ranked universities.

\begin{table}[t]
\centering
\caption{Aggregate evaluation metrics (THE, $n{=}352$).}
\label{tab:aggregate}
\small
\begin{tabular}{@{}lr@{}}
\toprule
\textbf{Metric} & \textbf{Value} \\
\midrule
MAE & 251.5 \\
Median AE & 131.5 \\
RMSE & 411.4 \\
PNMAE & 12.03\% \\
Signed Error & $+190.1$ \\
Spearman's $\rho$ & 0.769 \\
Pearson's $r$ & 0.677 \\
Kendall's $\tau$ & 0.591 \\
\midrule
Hit Rate @25 & 10.2\% (36/352) \\
Hit Rate @50 & 20.7\% (73/352) \\
Hit Rate @100 & 39.8\% (140/352) \\
Wilson-claimed @50 & $\geq$16\% \\
\midrule
Range Coverage & 8.2\% (29/352) \\
Mean Range Width & 42.9 positions \\
\midrule
Calibration Slope $\beta$ & 0.951 \\
Calibration Intercept $\alpha$ & 209.9 \\
Cohen's $\kappa$ (tier) & 0.349 (fair) \\
Tier Agreement & 49.7\% (175/352) \\
\midrule
Memorization Index & 0.000 (0/352) \\
AE${=}0$ (exact matches) & 1 (Northwestern U.) \\
\bottomrule
\end{tabular}
\end{table}

The positive signed error ($+190.1$) indicates the system systematically predicts \emph{worse} ranks (higher numbers) than actual. This is expected: bibliometrics alone cannot capture reputation and teaching signals, which tend to improve a university's actual rank beyond what research metrics predict.

\subsection{Per-Tier Performance}

Table~\ref{tab:pertier} breaks down performance by evaluation tier. MI${}=0.000$ across all tiers.

\begin{table}[t]
\centering
\caption{Per-tier performance breakdown (THE).}
\label{tab:pertier}
\small
\begin{tabular}{@{}lrrrrrr@{}}
\toprule
\textbf{Tier} & \textbf{N} & \textbf{MAE} & \textbf{Med.} & \textbf{@50} & \textbf{@100} & \textbf{SE} \\
\midrule
Elite & 21 & 60.5 & 33.0 & 57.1 & 90.5 & $+$60.2 \\
Strong & 80 & 211.6 & 93.0 & 30.0 & 53.8 & $+$205.8 \\
Mid & 99 & 247.8 & 122.5 & 18.2 & 38.4 & $+$214.0 \\
Lower & 99 & 287.0 & 192.0 & 11.1 & 29.3 & $+$198.5 \\
Tail & 53 & 328.2 & 309.0 & 15.1 & 20.8 & $+$157.2 \\
\bottomrule
\end{tabular}
\end{table}

Performance degrades monotonically from elite to tail, consistent with decreasing bibliometric signal density for lower-ranked universities. Elite tier performance (57.1\% hit@50, 90.5\% hit@100) contradicts what memorization would predict: if the LLM were memorizing, elite (most famous) universities would be recalled perfectly (AE$\approx$0), not with MAE${=}60.5$.

\begin{figure*}[t]
  \centering
  \includegraphics[width=\textwidth]{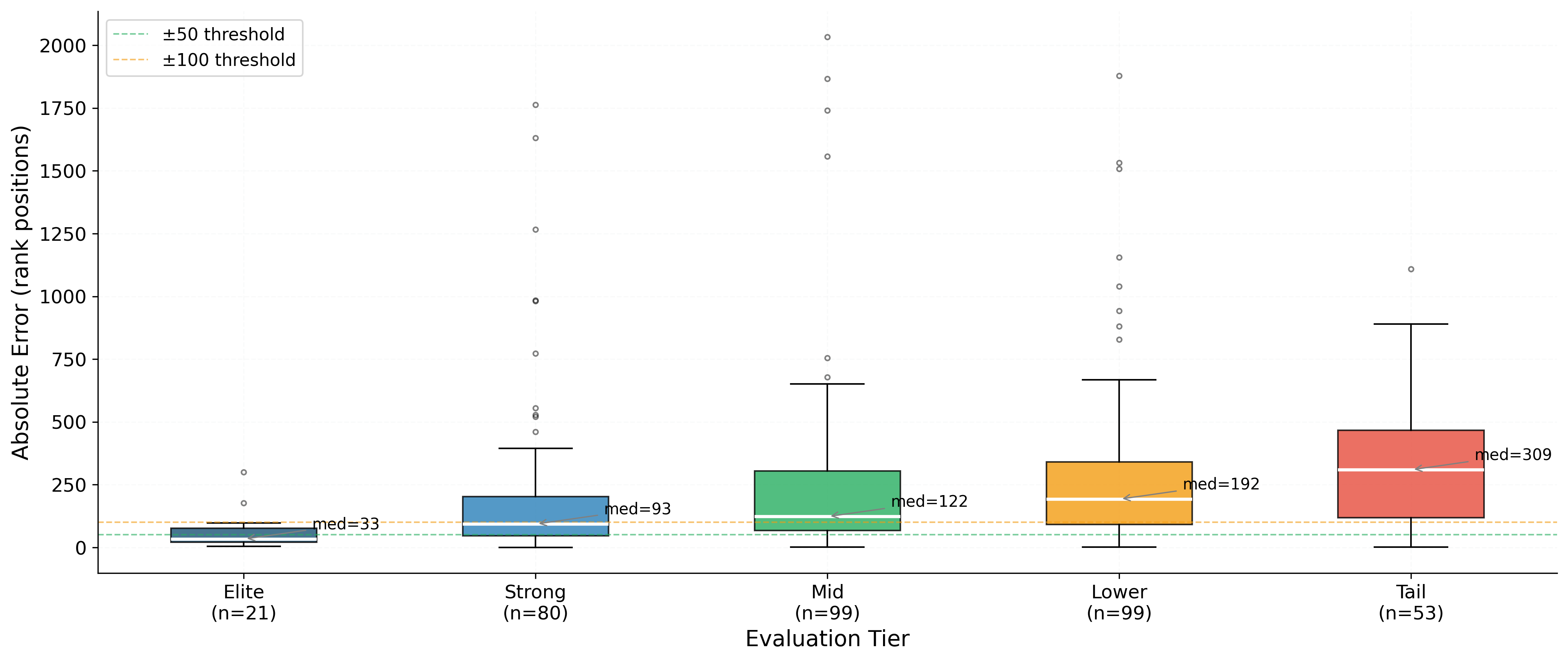}
  \caption{Error distribution by evaluation tier (THE). Box plots show median (line), IQR (box), and outliers. Performance degrades monotonically from elite to tail.}
  \label{fig:error_by_tier}
\end{figure*}

\begin{figure*}[t]
  \centering
  \includegraphics[width=\textwidth]{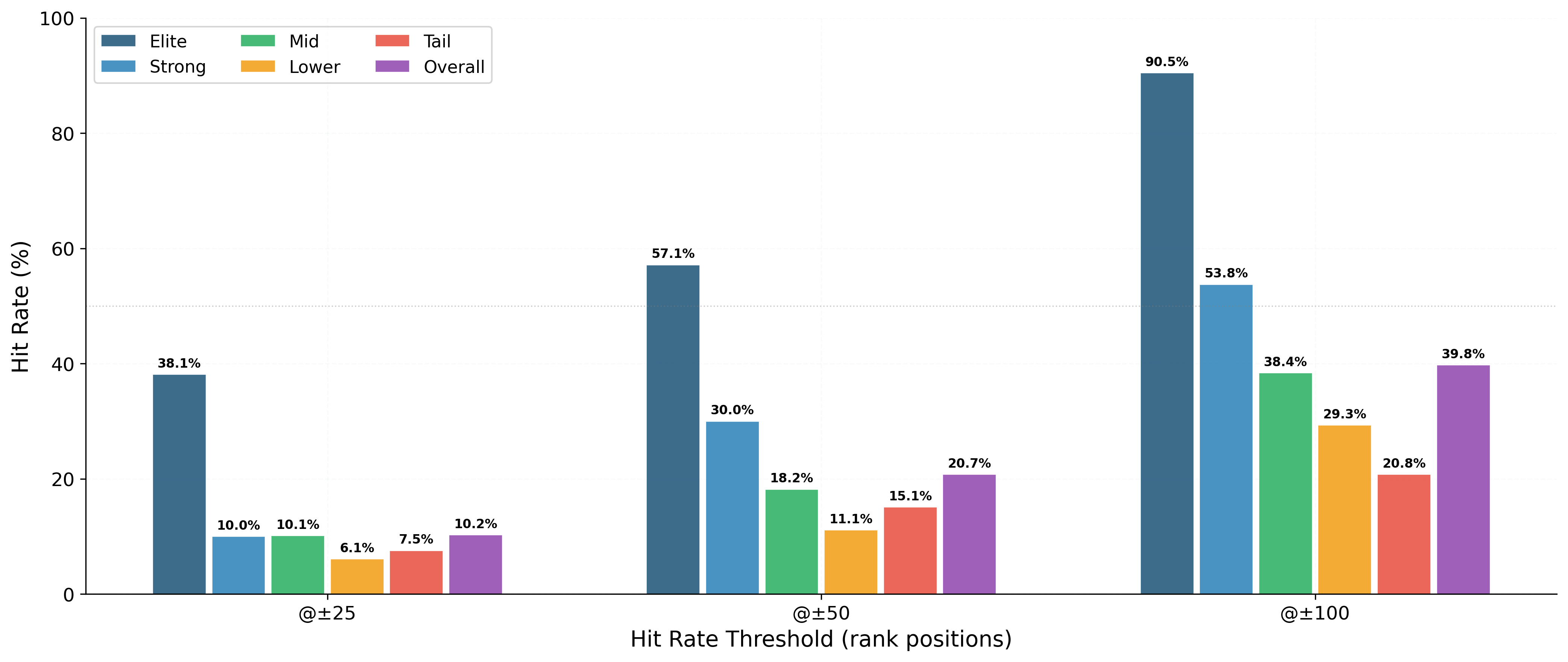}
  \caption{Hit rate comparison across tiers at three thresholds (@25, @50, @100). Elite tier achieves 90.5\% hit@100.}
  \label{fig:hit_rates}
\end{figure*}

Figure~\ref{fig:error_by_tier} shows the error distribution by tier, and Figure~\ref{fig:hit_rates} compares hit rates across tiers and thresholds.

\textbf{Tier confusion matrix.} Table~\ref{tab:confusion} shows actual vs.\ predicted tier classification.

\begin{table}[t]
\centering
\caption{Tier confusion matrix (actual $\times$ predicted).}
\label{tab:confusion}
\small
\begin{tabular}{@{}lccccc@{}}
\toprule
& \multicolumn{5}{c}{\textbf{Predicted}} \\
\cmidrule{2-6}
\textbf{Actual} & Elite & Strong & Mid & Lower & Tail \\
\midrule
Elite & 4 & 15 & 2 & 0 & 0 \\
Strong & 0 & 36 & 32 & 6 & 6 \\
Mid & 0 & 6 & 49 & 26 & 18 \\
Lower & 0 & 1 & 13 & 40 & 45 \\
Tail & 0 & 0 & 2 & 5 & 46 \\
\bottomrule
\end{tabular}
\end{table}

\subsection{Predicted vs.\ Actual Rank Correlation}

\begin{figure*}[t]
  \centering
  \includegraphics[width=\textwidth]{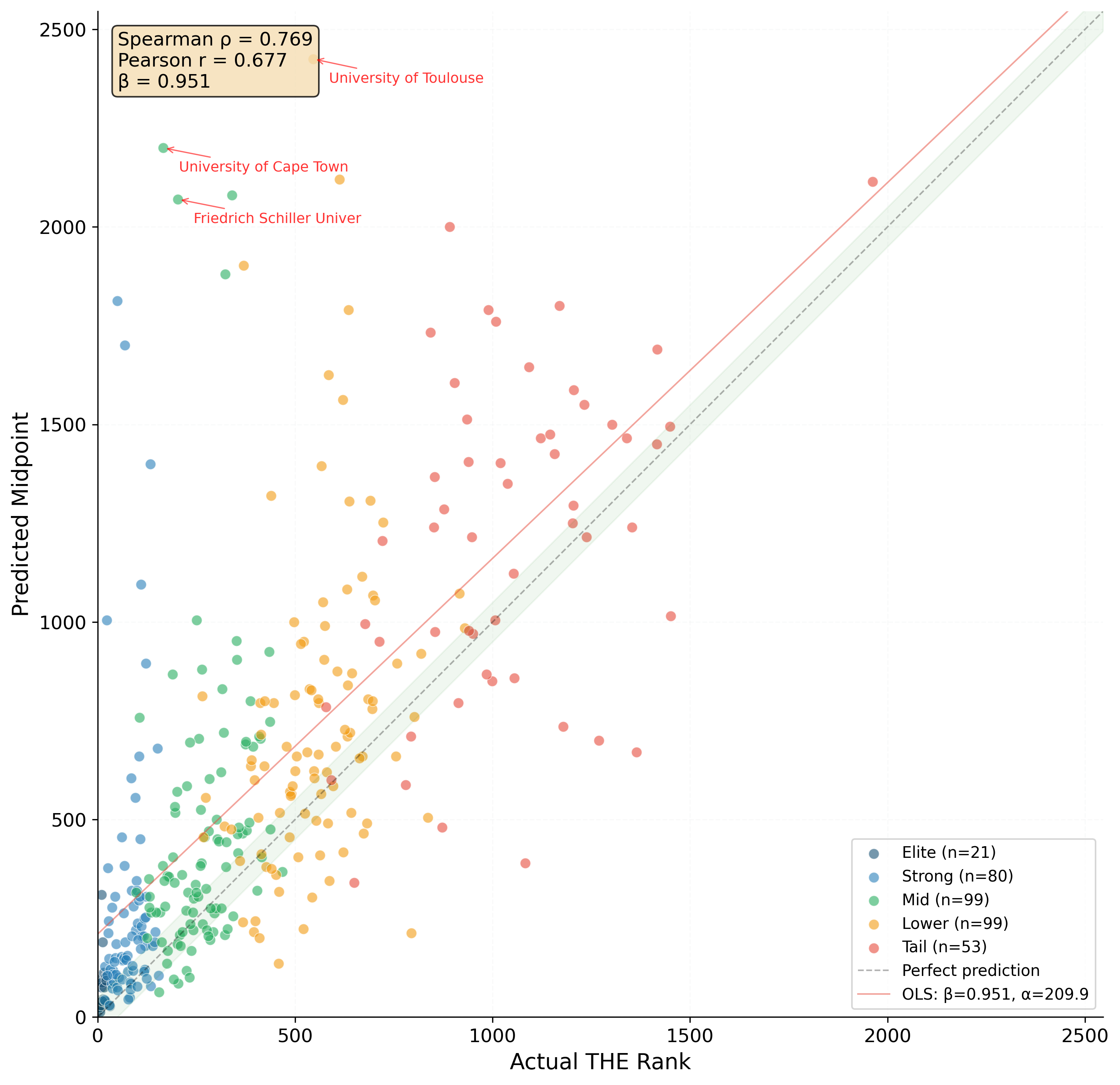}
  \caption{Predicted vs.\ actual rank (THE, $n{=}352$). Diagonal = perfect prediction; shaded band = $\pm 50$ positions. Spearman $\rho{=}0.769$, calibration slope $\beta{=}0.951$.}
  \label{fig:scatter}
\end{figure*}

Figure~\ref{fig:scatter} plots predicted midpoint against actual rank for all 352 universities, colored by tier. The Spearman correlation ($\rho{=}0.769$) and calibration slope ($\beta{=}0.951$) indicate meaningful ordinal agreement with remarkably low systematic compression bias.

\subsection{Failure Taxonomy}

All failures share a common root cause: the system estimates rankings using research quality proxies alone, while THE incorporates industry impact (4\%), reputation surveys (${\sim}18\%$), and teaching quality (${\sim}29.5\%$) that are entirely unobservable to our pipeline. Table~\ref{tab:failure_modes} defines six failure modes; Table~\ref{tab:failure_prevalence} summarizes their prevalence among the 20 worst predictions.

\begin{table*}[t]
\centering
\caption{Failure mode definitions (left) and prevalence in top-20 worst predictions (right).}
\label{tab:failure_modes}
\small
\begin{minipage}[t]{0.48\textwidth}
\centering
\begin{tabular}{@{}clp{4.8cm}@{}}
\toprule
\textbf{ID} & \textbf{Mode} & \textbf{Description} \\
\midrule
F1 & Reputation & Rank boosted by reputation/teaching signals unobservable from bibliometrics \\
F2 & Teaching & Rank driven by teaching metrics (student--staff ratio, doctoral ratio) \\
F3 & Scale & Small-but-excellent or large-but-average confusion \\
F4 & Regional & Systematic regional over/under-estimation \\
F5 & Anchor Drift & Calibration anchors on unrepresentative comparators \\
F6 & Data Gap & OpenAlex provides fragmented or incorrect institutional data \\
\bottomrule
\end{tabular}
\end{minipage}%
\hfill
\begin{minipage}[t]{0.48\textwidth}
\centering
\label{tab:failure_prevalence}
\begin{tabular}{@{}lrrl@{}}
\toprule
\textbf{Mode} & \textbf{Primary} & \textbf{\%} & \textbf{Key pattern} \\
\midrule
F1: Reputation & 13 & 65\% & Strong brand, weak metrics \\
F6: Data Gap & 7 & 35\% & OpenAlex inconsistency \\
F5: Anchor Drift & 5 (2nd) & 25\% & Amplifies F1/F6 \\
F4: Regional & 5 (2nd) & 25\% & Russian, French institutions \\
F3: Scale & 3 (2nd) & 15\% & Small specialized schools \\
\bottomrule
\end{tabular}
\end{minipage}
\end{table*}

\subsection{Case Studies}

We present three representative case studies spanning distinct failure modes.

\textbf{University of Toulouse (F6, AE${=}1{,}879$).} OpenAlex fragments this multi-campus system into separate entities; the pipeline retrieved data for only Universit\'{e} Toulouse III -- Paul Sabatier, yielding 2yr citedness${=}0.09$ and h-index${=}18$---catastrophically unrepresentative. The LLM's reasoning was sound given its inputs; the inputs were wrong.

\textbf{Yale University (F1, AE${=}300$).} Yale's 2yr citedness of 1.87 places it well below institutions like MIT (5.74) and Caltech (4.91) on pure research impact. The LLM correctly estimated ${\sim}290{-}330$ from bibliometrics alone, but Yale's THE rank of \#10 is driven substantially by reputation---one of the most recognized university brands globally. This error is entirely attributable to unobservable reputation signals.

\textbf{University of Michigan (F6+F1+F5, AE${=}982$).} Research excellence of 5.0\% and international collaboration of 8.1\% are implausibly low for THE \#23. The initial estimate (${\sim}200{-}500$) was already compromised by bad input data (F6). Calibration then compared against \#360--\#1250 universities, found metric matches around \#1000, and drifted the estimate to 980--1030 (F5), amplifying the error from ${\sim}350$ to 982---a textbook case of cascading failure.

\subsection{Statistical Test Results}

Table~\ref{tab:stats} reports statistical comparisons between initial (Stage~1) and calibrated (Stage~1+2) predictions.

\begin{table}[t]
\centering
\caption{Statistical test results (initial vs.\ calibrated).}
\label{tab:stats}
\small
\begin{tabular}{@{}lrrl@{}}
\toprule
\textbf{Test} & \textbf{Stat.} & \textbf{$p$} & \textbf{Interpretation} \\
\midrule
Wilcoxon & $W{=}27{,}921$ & 0.343 & Not significant \\
McNemar & $\chi^2{=}1.786$ & 0.181 & Not significant \\
Kruskal--Wallis & $H{=}42.09$ & $10^{-8}$ & Tiers differ$^{***}$ \\
\bottomrule
\end{tabular}
\end{table}

\textbf{Calibration impact.} Table~\ref{tab:calibration} summarizes the effect of calibration. Overall, calibration produces a marginal improvement (MAE: $256.8 \to 251.5$, $-2.1\%$) that is not statistically significant ($p{=}0.343$). However, the effect is bimodal: calibration reliably improves elite predictions (85.7\% improved) and tail predictions (56.6\% improved), but has mixed effects on mid-tier universities.

\begin{table}[t]
\centering
\caption{Calibration impact summary.}
\label{tab:calibration}
\small
\begin{tabular}{@{}lrrr@{}}
\toprule
\textbf{Metric} & \textbf{Initial} & \textbf{Calibrated} & \textbf{$\Delta$} \\
\midrule
MAE & 256.8 & 251.5 & $-2.1\%$ \\
Median AE & 149.5 & 130.8 & $-12.5\%$ \\
Spearman $\rho$ & 0.770 & 0.769 & $-0.001$ \\
Hit @50 & 17.9\% & 20.7\% & $+2.8$ pp \\
Hit @100 & 36.1\% & 39.8\% & $+3.7$ pp \\
\midrule
Improved & \multicolumn{3}{c}{191/352 (54.3\%)} \\
Worsened & \multicolumn{3}{c}{153/352 (43.5\%)} \\
Unchanged & \multicolumn{3}{c}{8/352 (2.3\%)} \\
\bottomrule
\end{tabular}
\end{table}

\begin{figure*}[t]
  \centering
  \includegraphics[width=\textwidth]{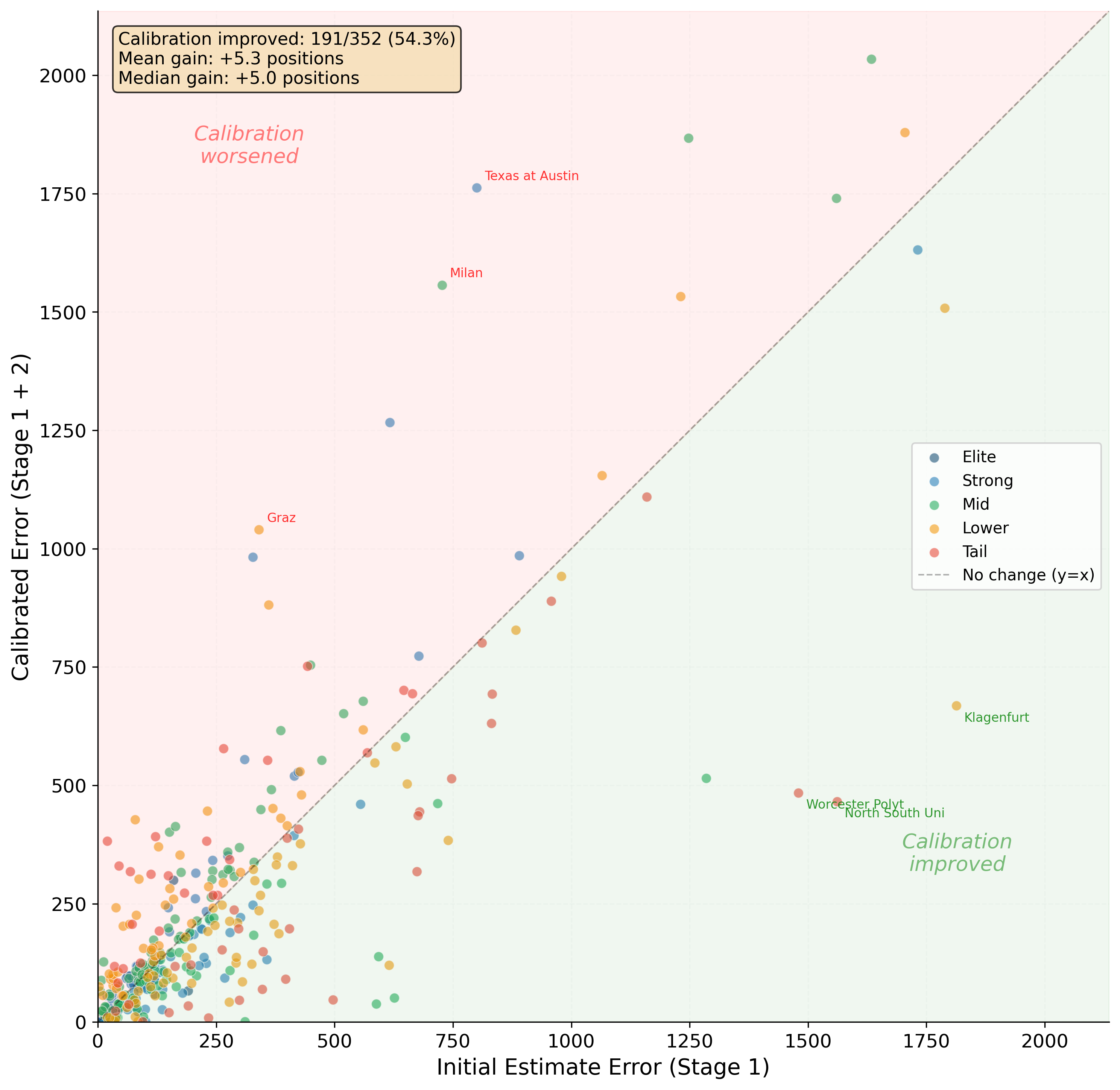}
  \caption{Initial vs.\ calibrated absolute error. Points below the diagonal indicate calibration helped. Calibration improved 54.3\% of predictions overall.}
  \label{fig:calibration_gain}
\end{figure*}

Figure~\ref{fig:calibration_gain} visualizes the per-university calibration effect.

\section{Discussion}\label{sec:discussion}

\subsection{Beyond Memorization: Evidence and Arguments}

We structure the anti-memorization argument across six lines of evidence:

\textbf{1.~Anonymization prevents name-based recall.} The LLM receives \texttt{INST-A3F2B1C4} with numeric metrics only. Even if a metric profile seems familiar, the model must reason about rank implications---not retrieve a memorized answer.

\textbf{2.~Data hiding prevents tool-based leakage.} The \texttt{get\_ranking\_samples} tool physically cannot return the target university's rank.

\textbf{3.~Calibration requires reasoning, not retrieval.} The multi-step process of fetching samples, computing metrics, comparing, and determining relative positioning is an analytical workflow, not a lookup.

\textbf{4.~Large errors on famous universities.} The system produces its largest errors on well-known institutions---the opposite of what memorization would predict. Yale (THE \#10, predicted 310, AE${=}300$), Oxford (THE \#1, predicted 95, AE${=}94$), Michigan (THE \#23, predicted 1005, AE${=}982$), and UT Austin (THE \#50, predicted 1813, AE${=}1{,}763$) are among the most-discussed universities in training corpora. A memorizing system would recall these perfectly.

\textbf{5.~Memorization Index.} MI${=}0.000$ (0/352). Zero universities received an exact point prediction with a zero-width range. The single AE${=}0$ case (Northwestern University) had a range width of 55 positions ([328, 383], actual${=}355$)---a fortunate analytical prediction, not memorized retrieval.

\textbf{6.~Systematic positive signed error.} The system consistently predicts worse ranks than actual ($+190.1$) across all tiers, consistent with missing reputation/teaching signals rather than memorized recall (which would produce signed error $\approx 0$).

\subsection{Observable vs.\ Unobservable Data}

\begin{figure*}[t]
  \centering
  \includegraphics[width=\textwidth]{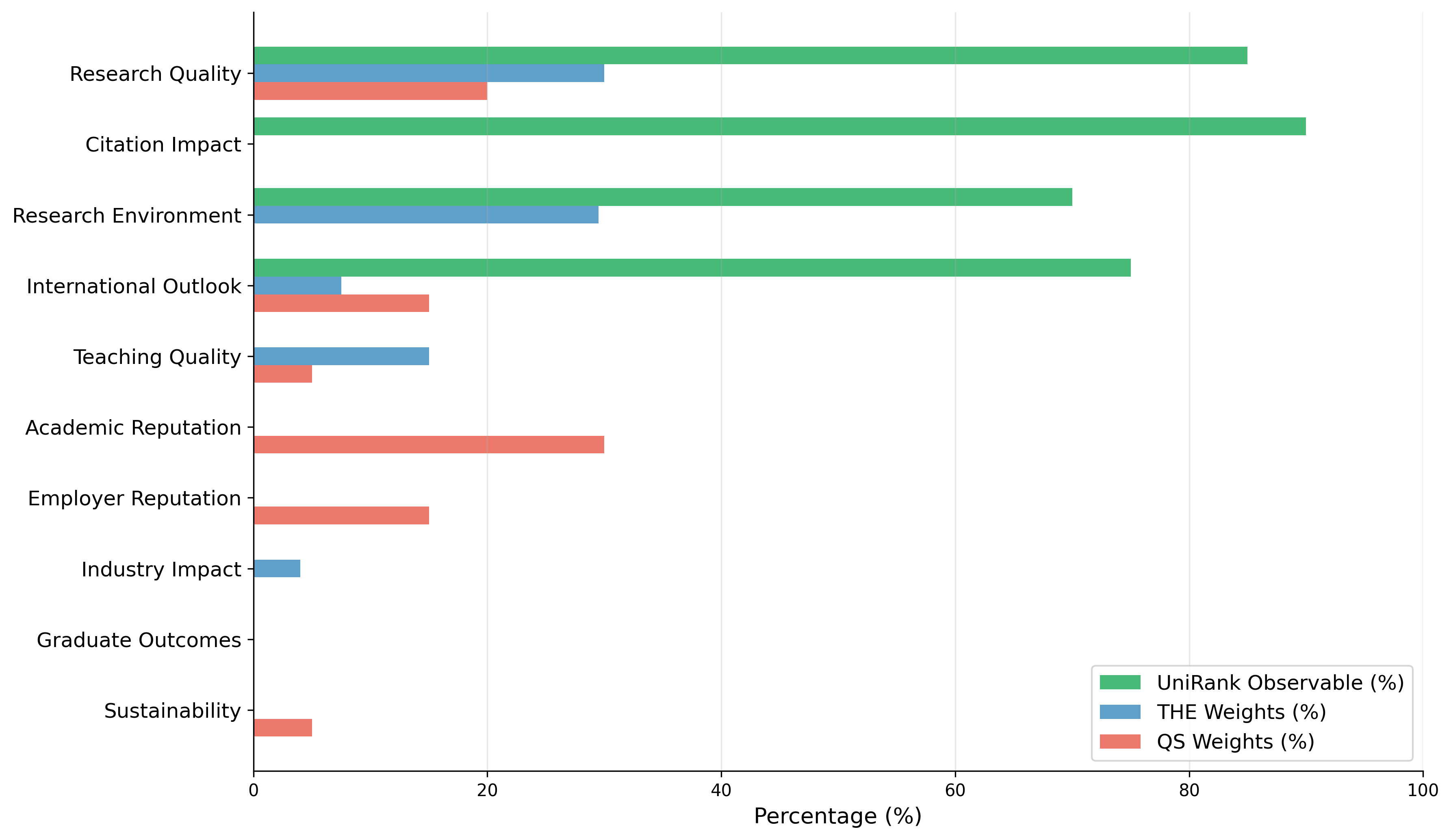}
  \caption{Data completeness radar. Green = what UniRank can measure; gray = what THE weights. The gap between series---particularly on Teaching (15\%), Reputation (18\%), and Industry (4\%)---explains the systematic positive signed error.}
  \label{fig:radar}
\end{figure*}

Figure~\ref{fig:radar} illustrates the fundamental challenge: UniRank can observe research metrics well but has zero signal on reputation, teaching, and industry dimensions.

Table~\ref{tab:data_gap} details the missing data dimensions and their impact.

\begin{table}[t]
\centering
\caption{Missing data dimensions and ranking impact.}
\label{tab:data_gap}
\small
\begin{tabular}{@{}lll@{}}
\toprule
\textbf{Missing Data} & \textbf{Impact} & \textbf{Systems} \\
\midrule
Reputation surveys & 30\% QS, 18\% THE & QS, THE \\
Employer surveys & 15\% QS & QS \\
Student--staff ratios & 5\% QS, 15\% THE & QS, THE \\
Intl.\ student/faculty & 10\% QS & QS \\
Nobel/Fields winners & 30\% ARWU & ARWU \\
Industry income & 4\% THE & THE \\
\bottomrule
\end{tabular}
\end{table}

\subsection{Limitations}

\begin{enumerate}[nosep]
  \item \textbf{Sample size}: 352 successful predictions from THE (${\sim}17\%$ of 2{,}092 ranked universities).
  \item \textbf{Single system}: Results are for THE only; QS and ARWU evaluations remain future work.
  \item \textbf{Model dependency}: Results are specific to GPT-5.2; different models may yield different results.
  \item \textbf{Temporal validity}: Rankings and bibliometric data evolve annually.
  \item \textbf{Cost}: Each evaluation requires 4--6 LLM calls with up to 12 tool-call steps each.
  \item \textbf{Single-run results}: Due to LLM non-determinism, results may vary across runs. We report single-run results without repeated trials.
\end{enumerate}

\section{Future Work}\label{sec:future}

\begin{enumerate}[nosep]
  \item \textbf{Multi-system evaluation}: Run stratified test sets across QS and ARWU.
  \item \textbf{Additional data sources}: Student/staff data from government databases, patent data from Lens.org, Nobel/Fields data for ARWU, webometrics.
  \item \textbf{LLM-driven reputation gathering}: Use agentic web search and deep-research workflows to let the LLM retrieve real-time reputation signals---news coverage, employer surveys, faculty awards.
  \item \textbf{Adaptive calibration}: Select comparison universities by metric similarity rather than rank proximity.
  \item \textbf{TOPSIS-based aggregation}: Integrate the Technique for Order of Preference by Similarity to Ideal Solution (TOPSIS) as a complementary multi-criteria decision-making layer, leveraging its ability to rank alternatives by geometric distance from ideal and anti-ideal profiles across heterogeneous indicator dimensions---providing a mathematically grounded aggregation baseline against which the LLM-based calibration can be compared and combined.
\end{enumerate}

\section{Conclusion}\label{sec:conclusion}

We presented UniRank, a multi-agent calibration pipeline that estimates university rankings from publicly available bibliometric data, using anonymization and data hiding to isolate LLM reasoning from training data memorization.

On the THE World University Rankings ($n{=}352$), UniRank achieves MAE${=}251.5$ rank positions, Spearman~$\rho{=}0.769$, hit rate @100${=}39.8\%$, and a Memorization Index of exactly zero. Performance is strongest in the elite tier (MAE${=}60.5$, hit@100${=}90.5\%$) and degrades predictably toward the tail (MAE${=}328.2$, hit@100${=}20.8\%$). The systematic positive signed error ($+190.1$) across all tiers confirms the system's core limitation: it cannot observe reputation and teaching signals that account for approximately 50\% of the THE ranking methodology. Despite this, the correlation metrics ($\rho{=}0.769$, $\tau{=}0.591$, $r{=}0.677$) demonstrate meaningful ordinal agreement, and the calibration slope ($\beta{=}0.951$) shows remarkably low systematic compression bias.

Our failure analysis reveals two dominant error sources: \textbf{F1 (Reputation Blind Spot)}---accounting for 65\% of the top-20 worst predictions---and \textbf{F6 (Data Source Incompleteness)}---35\% of the top-20 worst.

\textbf{The primary contribution of this work is not a production ranking system, but an evaluation framework.} UniRank establishes a rigorous, reproducible methodology---including the leave-one-out protocol, anonymization-based decontamination, memorization index measurement, and multi-dimensional failure taxonomy---that enables systematic assessment of future improvements. As we incorporate more complete data sources for reputation, teaching quality, and graduate outcomes, this evaluation framework will measure whether each addition genuinely improves ranking estimation or merely adds noise. The current results (Spearman~$\rho{=}0.769$ from research proxies alone) provide a clear baseline against which all future enhancements can be benchmarked.

A live demo is available at \url{https://unirank.scinito.ai}.

\section*{Acknowledgments}

We thank OpenAlex and Semantic Scholar for providing open access to scholarly metadata, and the QS, THE, and ARWU organizations for publishing their ranking methodologies.

\bibliographystyle{plainnat}
\bibliography{references}

\appendix
\onecolumn

\section{Reproducibility Checklist}\label{app:repro}

\begin{table}[H]
\centering
\begin{tabular}{@{}ll@{}}
\toprule
\textbf{Item} & \textbf{Detail} \\
\midrule
Model & GPT-5.2 (via \texttt{@ai-sdk/openai}, Helicone gateway) \\
Reasoning effort & \texttt{medium} \\
Temperature & Default (model-determined) \\
Max calibration steps & 12 per system \\
Structured output & Zod schemas via Vercel AI SDK \\
Eval concurrency & 3 parallel evaluations \\
Eval delay & 2000ms between batches \\
Non-determinism & Single-run results; may vary across runs \\
Data snapshot & OpenAlex Feb 2026; QS 2026, THE 2024, ARWU 2025 \\
Code availability & Live demo at \url{https://unirank.scinito.ai} \\
\bottomrule
\end{tabular}
\end{table}

\section{Full Tier Benchmark Table}\label{app:benchmarks}

Table~\ref{tab:benchmarks} shows the empirical benchmarks derived from OpenAlex data for four tiers.

\begin{table}[H]
\centering
\caption{Tier benchmarks (mean $\pm$ SD) for key metrics.}
\label{tab:benchmarks}
\small
\begin{tabular}{@{}lcccc@{}}
\toprule
\textbf{Metric} & \textbf{Tier 1 (Top 10)} & \textbf{Tier 2 (Top 100)} & \textbf{Tier 3 (Top 300)} & \textbf{Tier 4 (Top 500)} \\
\midrule
h-index & 2135 $\pm$ 291 & 1196 $\pm$ 205 & 787 $\pm$ 145 & 560 $\pm$ 102 \\
2yr citedness & 4.60 $\pm$ 1.10 & 3.40 $\pm$ 0.70 & 2.95 $\pm$ 0.60 & 2.85 $\pm$ 0.77 \\
Research excellence (\%) & 19.5 $\pm$ 3.5 & 20.5 $\pm$ 2.1 & 20.0 $\pm$ 2.8 & 19.5 $\pm$ 3.5 \\
Intl.\ collab (\%) & 40.5 $\pm$ 9.2 & 40.5 $\pm$ 12.0 & 39.0 $\pm$ 12.7 & 38.0 $\pm$ 12.7 \\
\bottomrule
\end{tabular}
\end{table}

\begin{figure*}[htbp]
  \centering
  \includegraphics[width=\textwidth]{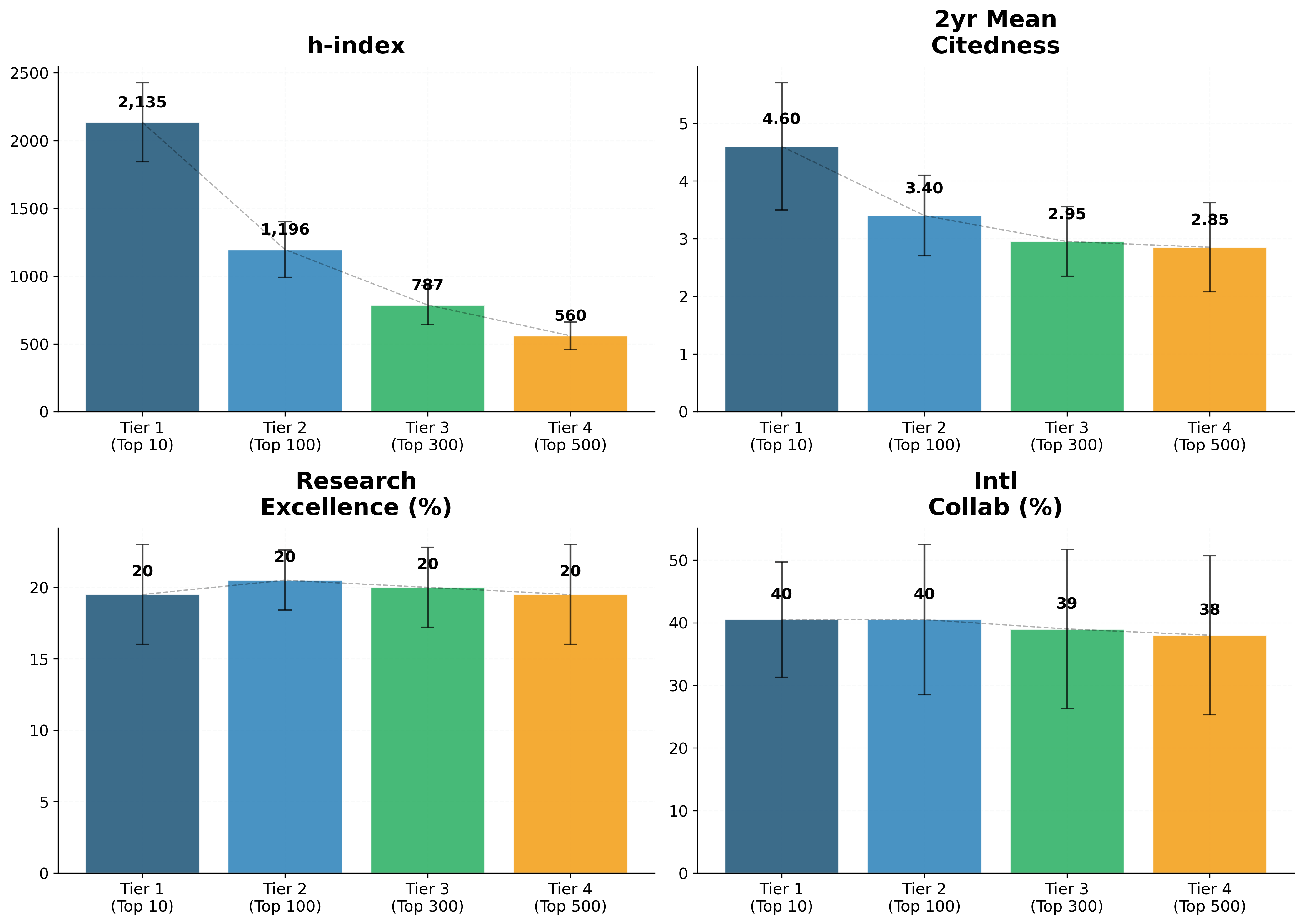}
  \caption{Tier benchmarks for key bibliometric indicators. Error bars show $\pm 1$ SD.}
  \label{fig:tier_benchmarks}
\end{figure*}

\section{Prompt Templates}\label{app:prompts}

\subsection{Stage 1: Initial Estimator}

\begin{lstlisting}[caption={Initial estimation prompt (abbreviated).}]
You are a university ranking estimation engine.

You receive anonymized bibliometric data about an institution. Your job
is to estimate where this institution would rank in QS, THE, and ARWU.

## What You Know About Each System

### QS 2025 (50% Research, 45% Reputation, 5% Sustainability)
- Academic Reputation (30%): Proxy from h-index, citation volume
- Citations per Faculty (20%): Citation metrics, NORMALIZED
- Employer Reputation (15%): Infer from institution type
- International Research Network (5%): International collaboration rate

### THE 2025 (30% Research Quality, 29.5% Env., 15% Teaching)
- Research Excellence (5%): % of works in top 10% by FWCI
- Citation Impact (15%): FWCI-based
- Research Reputation (18%): Infer from h-index + citation volume
- International Outlook (7.5%): Collaboration %

### ARWU (100% Hard Research Metrics, no reputation)
- Highly Cited Researchers (20%): h-index as proxy
- Nature/Science papers (20%): Top venue publications
- Total publications (20%): Works count
- Nobel/Fields (30%): Not in data

## Rules
- Estimating based on METRICS ONLY
- All data is anonymized -- do NOT try to identify the institution
- Produce a RANGE (min-max), never a point estimate
- Include confidence: "high" (+/-20), "medium" (+/-50), "low" (+/-100+)
\end{lstlisting}

\subsection{Stage 2: THE Calibration Engine}

\begin{lstlisting}[caption={THE calibration prompt (abbreviated).}]
You are a THE World University Rankings calibration engine.

You receive anonymized bibliometric metrics and an initial rank estimate.
Refine this estimate by comparing against real universities in THE.

## THE Ranking Weights
- Research Quality (30%): Research Excellence %, FWCI, influential cites
- Research Environment (29.5%): h-index, productivity, reputation
- Teaching (15%): Not directly measurable
- International Outlook (7.5%): International collaboration %
- Industry Impact (4%): Limited data

## Calibration Protocol
1. Start with the initial estimate range
2. Call get_ranking_samples to fetch 3 universities near estimate center
3. Call compute_metrics for 2-3 sample universities
4. Compare target's metrics against each sample
5. Decide if target ranks ABOVE, BETWEEN, or BELOW samples
6. Adjust range; MAY move outside initial estimate if warranted
7. Produce final calibrated range (target width: 30-50 positions)

## Rules
- ALWAYS use tools; budget of 10 tool-call steps
- Aim for range width of 30-50 positions
- DO NOT try to identify the target institution
\end{lstlisting}

\subsection{Stage 3: Final Report Synthesis}

\begin{lstlisting}[caption={Final report prompt (abbreviated).}]
You are UniRank, an expert analyst in global higher education rankings.

Writing the FINAL REPORT for a university analysis. The rank estimation
has been done; you receive the results. Synthesize into a structured
analysis.

## What You Receive
1. Full university profile with bibliometric data (unblinded)
2. Initial zero-shot estimate (Stage 1)
3. Calibration results from per-system engines (Stage 2)
4. Published rankings for systems where university is already ranked

## Output Format
- Estimated Global Rank Range
- Methodology Breakdown (per system)
- Strengths (3-5, each referencing a metric)
- Weaknesses (3-5, each referencing a metric or missing data)
- Citation Quality Assessment (when S2 data available)
- Comparable Peers (3-5 universities)
- Strategic Recommendations (2-3 actionable suggestions)

## Rules
- NEVER invent specific ranking numbers
- ALWAYS cite specific data values
- When data is MISSING, explicitly state this
\end{lstlisting}

\section{Test Set Sampling Algorithm}\label{app:sampling}

\begin{algorithm}[H]
\caption{Stratified Test Set Construction}
\label{alg:sampling}
\KwIn{Universe $\mathcal{U}$ of universities with OpenAlex IDs; target size $N$}
\KwOut{Stratified test set $\mathcal{T}$}
\BlankLine
\tcp{Compute consensus rank for each university}
\ForEach{$u \in \mathcal{U}$}{
  $r_u \gets \text{median}(\{r_s(u) : s \in \{QS, THE, ARWU\}, r_s(u) \neq \text{null}\})$\;
}
\BlankLine
\tcp{Assign tiers}
$\text{tiers} \gets \{$Elite: $[1,25]$, Strong: $[26,150]$, Mid: $[151,400]$, Lower: $[401,700]$, Tail: $[700+]\}$\;
\ForEach{$u \in \mathcal{U}$}{
  $\text{tier}(u) \gets$ tier containing $r_u$\;
}
\BlankLine
\tcp{Assign regions from country code}
$\text{regions} \gets \{$NA, Europe, Asia-Pacific, Rest of World, Mixed$\}$\;
\ForEach{$u \in \mathcal{U}$}{
  $\text{region}(u) \gets \text{mapCountryToRegion}(\text{country}(u))$\;
}
\BlankLine
\tcp{Stratified sampling: equal tier representation}
$\mathcal{T} \gets \emptyset$\;
$n_{\text{per\_tier}} \gets N / |\text{tiers}|$\;
\ForEach{$t \in \text{tiers}$}{
  $\mathcal{U}_t \gets \{u : \text{tier}(u) = t\}$\;
  $\mathcal{T} \gets \mathcal{T} \cup \text{sampleProportional}(\mathcal{U}_t, n_{\text{per\_tier}}, \text{by=region})$\;
}
\Return{$\mathcal{T}$}
\end{algorithm}

\section{Full Results Table}\label{app:results}

Table~\ref{tab:full_results} presents all 352 successful predictions sorted by absolute error (ascending). Best prediction: Northwestern University (AE${=}0$). Worst prediction: University of Cape Town (AE${=}2{,}034$).

{\small
\begin{longtable}{@{}rlrrrl@{}}
\caption{Full evaluation results (THE, $n{=}352$), sorted by AE ascending.} \label{tab:full_results} \\
\toprule
\textbf{\#} & \textbf{University} & \textbf{Actual} & \textbf{Pred.} & \textbf{AE} & \textbf{Tier} \\
\midrule
\endfirsthead
\multicolumn{6}{c}{\tablename\ \thetable\ -- continued} \\
\toprule
\textbf{\#} & \textbf{University} & \textbf{Actual} & \textbf{Pred.} & \textbf{AE} & \textbf{Tier} \\
\midrule
\endhead
\midrule
\multicolumn{6}{r}{Continued on next page} \\
\endfoot
\bottomrule
\endlastfoot
1 & Northwestern University & 30 & 30 & 0 & strong \\
2 & Tianjin University & 208 & 208 & 1 & mid \\
3 & Univ.\ of Kwazulu-Natal & 566 & 565 & 1 & lower \\
4 & Hong Kong Baptist Univ. & 214 & 215 & 1 & mid \\
5 & Vrije Univ.\ Brussel & 234 & 235 & 1 & mid \\
6 & Univ.\ of Alberta & 119 & 120 & 1 & strong \\
7 & Univ.\ of Glasgow & 84 & 85 & 1 & strong \\
8 & Lithuanian U.\ Health Sci. & 1006 & 1005 & 1 & tail \\
9 & Brunel Univ.\ London & 414 & 413 & 2 & lower \\
10 & Univ.\ of Maryland & 117 & 115 & 2 & strong \\
11 & Nanyang Technological U. & 31 & 28 & 4 & elite \\
12 & Hong Kong Polytechnic U. & 82 & 88 & 6 & strong \\
13 & Harvard University & 6 & 13 & 7 & elite \\
14 & Univ.\ of Minho & 663 & 655 & 8 & lower \\
15 & New Jersey Inst.\ Tech. & 591 & 600 & 9 & tail \\
16 & Durham University & 177 & 168 & 10 & mid \\
17 & Univ.\ of Porto & 415 & 405 & 10 & mid \\
18 & U.\ Polit\`{e}cnica Catalunya & 670 & 660 & 10 & lower \\
19 & Univ.\ of Hull & 525 & 515 & 10 & lower \\
20 & Gachon University & 596 & 585 & 11 & lower \\
21 & East China Normal Univ. & 286 & 275 & 11 & mid \\
22 & Univ.\ of Groningen & 83 & 70 & 13 & strong \\
23 & UCL & 22 & 35 & 13 & elite \\
24 & MIT & 2 & 16 & 14 & elite \\
25 & Stanford University & 5 & 20 & 15 & elite \\
26 & Univ.\ of Sci.\ \& Tech.\ China & 51 & 68 & 17 & strong \\
27 & Univ.\ of Calgary & 202 & 185 & 17 & mid \\
28 & Old Dominion University & 950 & 970 & 20 & tail \\
29 & Princeton University & 3 & 24 & 21 & elite \\
30 & Cornell University & 19 & 40 & 21 & elite \\
31 & Imperial College London & 8 & 30 & 22 & elite \\
32 & Univ.\ of Padua & 242 & 220 & 22 & mid \\
33 & Bharathidasan Univ. & 1238 & 1215 & 23 & tail \\
34 & Univ.\ of Birmingham & 100 & 78 & 23 & strong \\
35 & Heinrich Heine U.\ D\"{u}sseldorf & 299 & 275 & 24 & mid \\
36 & Link\"{o}ping University & 241 & 265 & 24 & mid \\
37 & Heidelberg University & 49 & 75 & 26 & strong \\
38 & NUS & 17 & 43 & 26 & strong \\
39 & Wuhan University & 124 & 98 & 27 & strong \\
40 & Univ.\ of Copenhagen & 90 & 118 & 28 & strong \\
41 & Tsinghua University & 12 & 40 & 28 & elite \\
42 & Southern U.\ Sci.\ \& Tech. & 162 & 190 & 28 & mid \\
43 & Queensland U.\ of Tech. & 210 & 180 & 30 & mid \\
44 & Bo\u{g}azi\c{c}i University & 486 & 455 & 31 & lower \\
45 & King Saud University & 266 & 235 & 31 & mid \\
46 & Univ.\ of Pennsylvania & 14 & 45 & 31 & elite \\
47 & Univ.\ of Queensland & 81 & 50 & 32 & strong \\
48 & Boston University & 77 & 45 & 32 & strong \\
49 & Caltech & 7 & 40 & 33 & elite \\
50 & Univ.\ of Amsterdam & 62 & 95 & 33 & strong \\
51 & Hanyang University & 295 & 263 & 33 & mid \\
52 & Zewail City & 1416 & 1450 & 34 & tail \\
53 & Univ.\ of Trento & 360 & 395 & 35 & lower \\
54 & Univ.\ of Melbourne & 37 & 73 & 36 & strong \\
55 & Halmstad University & 940 & 978 & 38 & tail \\
56 & Stony Brook University & 313 & 275 & 38 & mid \\
57 & U.\ Teknologi Malaysia & 437 & 475 & 38 & mid \\
58 & Univ.\ of Cambridge & 4 & 43 & 39 & elite \\
59 & Sorbonne University & 76 & 115 & 39 & strong \\
60 & Trinity College Dublin & 175 & 135 & 40 & mid \\
61 & Carleton University & 580 & 620 & 40 & lower \\
62 & Univ.\ of Freiburg & 139 & 180 & 41 & strong \\
63 & Univ.\ of Minnesota & 88 & 130 & 42 & strong \\
64 & Univ.\ of Delhi & 802 & 760 & 42 & lower \\
65 & Natl.\ Taiwan Ocean U. & 1449 & 1495 & 46 & tail \\
66 & U.\ Libre de Bruxelles & 224 & 270 & 46 & mid \\
67 & Univ.\ of Liverpool & 144 & 190 & 46 & strong \\
68 & Aston University & 427 & 380 & 47 & lower \\
69 & The New School & 1203 & 1250 & 47 & tail \\
70 & Univ.\ of Sydney & 53 & 100 & 47 & strong \\
71 & Western Australia & 154 & 105 & 49 & strong \\
72 & EPFL & 35 & 85 & 50 & strong \\
73 & Shanghai Jiao Tong U. & 40 & 90 & 50 & strong \\
74 & Kyung Hee University & 254 & 305 & 51 & mid \\
75 & NW Polytechnical U. & 274 & 325 & 51 & mid \\
76 & Univ.\ of Dundee & 325 & 380 & 55 & mid \\
77 & Adelaide University & 134 & 79 & 56 & strong \\
78 & Nanjing U.\ Info.\ Sci. & 929 & 985 & 56 & lower \\
79 & Missouri U.\ Sci.\ \& Tech. & 549 & 605 & 56 & lower \\
80 & Xiamen University & 276 & 220 & 56 & mid \\
81 & Univ.\ of Plymouth & 553 & 498 & 56 & lower \\
82 & Univ.\ of Waikato & 461 & 518 & 57 & lower \\
83 & Ulsan NIST & 243 & 300 & 57 & mid \\
84 & Johns Hopkins Univ. & 16 & 75 & 59 & elite \\
85 & Univ.\ of Florence & 355 & 415 & 60 & mid \\
86 & Univ.\ of Basel & 120 & 180 & 60 & strong \\
87 & KU Leuven & 46 & 108 & 62 & strong \\
88 & Univ.\ of Bern & 109 & 173 & 64 & strong \\
89 & Prince Sattam Bin Abdulaziz U. & 440 & 375 & 65 & lower \\
90 & UIC & 250 & 315 & 65 & mid \\
91 & UC Berkeley & 9 & 75 & 66 & elite \\
92 & Univ.\ of Nottingham & 146 & 215 & 69 & strong \\
93 & Univ.\ of Sussex & 237 & 168 & 70 & mid \\
94 & King's College London & 38 & 108 & 70 & strong \\
95 & Ming Chi U.\ of Tech. & 1053 & 1123 & 70 & tail \\
96 & Florida Intl.\ University & 489 & 560 & 71 & lower \\
97 & Univ.\ of Toronto & 21 & 93 & 72 & elite \\
98 & UC Santa Barbara & 72 & 145 & 73 & strong \\
99 & Univ.\ of Nebraska-Lincoln & 548 & 623 & 75 & lower \\
100 & Univ.\ of Tehran & 490 & 565 & 75 & lower \\
101 & South China U.\ Tech. & 280 & 205 & 75 & mid \\
102 & Univ.\ of Lausanne & 125 & 200 & 75 & mid \\
103 & ETH Zurich & 11 & 88 & 77 & elite \\
104 & Southeast University & 292 & 215 & 77 & mid \\
105 & Zhejiang University & 39 & 118 & 79 & strong \\
106 & Nanjing University & 63 & 143 & 80 & strong \\
107 & Univ.\ of S.\ California & 75 & 155 & 80 & strong \\
108 & Brown University & 65 & 145 & 80 & strong \\
109 & Univ.\ of Jordan & 639 & 720 & 81 & lower \\
110 & Carnegie Mellon Univ. & 24 & 105 & 81 & strong \\
111 & Lincoln University & 603 & 685 & 82 & lower \\
112 & Columbia University & 20 & 103 & 83 & elite \\
113 & Univ.\ of Derby & 793 & 710 & 83 & tail \\
114 & Jamia Millia Islamia & 487 & 570 & 83 & lower \\
115 & Washington U.\ St.\ Louis & 67 & 150 & 83 & strong \\
116 & Dalian U.\ of Tech. & 404 & 320 & 84 & mid \\
117 & Duy Tan University & 695 & 780 & 85 & lower \\
118 & Sichuan University & 229 & 315 & 86 & mid \\
119 & Medical U.\ of Graz & 248 & 335 & 87 & mid \\
120 & NYU & 32 & 120 & 88 & strong \\
121 & Technion & 343 & 255 & 88 & mid \\
122 & McMaster University & 116 & 205 & 89 & strong \\
123 & Curtin University & 285 & 195 & 90 & mid \\
124 & Tokyo U.\ Agri.\ \& Tech. & 1204 & 1295 & 91 & tail \\
125 & Beijing U.\ Chem.\ Tech. & 451 & 360 & 91 & lower \\
126 & Univ.\ of Cyprus & 493 & 585 & 92 & lower \\
127 & Radboud University & 155 & 63 & 92 & mid \\
128 & Monash University & 60 & 153 & 93 & strong \\
129 & Ewha Womans University & 583 & 490 & 93 & lower \\
130 & Emory University & 102 & 195 & 93 & strong \\
131 & Univ.\ of Massachusetts & 112 & 205 & 93 & strong \\
132 & Univ.\ of Oxford & 1 & 95 & 94 & elite \\
133 & Polytechnic U.\ Valencia & 755 & 660 & 95 & lower \\
134 & Univ.\ of Otago & 377 & 473 & 96 & mid \\
135 & Univ.\ Paris Cit\'{e} & 192 & 95 & 97 & mid \\
136 & Univ.\ of Chicago & 15 & 113 & 98 & elite \\
137 & UC Merced & 407 & 505 & 98 & lower \\
138 & Univ.\ College Cork & 367 & 465 & 98 & mid \\
139 & Georgia Tech & 42 & 140 & 98 & strong \\
140 & Oxford Brookes Univ. & 819 & 920 & 101 & lower \\
141 & Univ.\ of Turin & 468 & 368 & 101 & mid \\
142 & Tabriz U.\ Med.\ Sci. & 626 & 728 & 102 & lower \\
143 & Shanghai University & 508 & 405 & 103 & lower \\
144 & Harokopio U.\ Athens & 696 & 800 & 104 & lower \\
145 & Univ.\ of British Columbia & 45 & 150 & 105 & strong \\
146 & Konkuk University & 559 & 665 & 106 & lower \\
147 & UC Riverside & 329 & 223 & 107 & mid \\
148 & Univ.\ of Auckland & 158 & 265 & 107 & mid \\
149 & Sun Yat-sen University & 225 & 118 & 108 & mid \\
150 & Autonomous U.\ Madrid & 384 & 493 & 109 & mid \\
151 & Dalhousie University & 354 & 463 & 109 & mid \\
152 & Univ.\ of Virginia & 171 & 280 & 109 & mid \\
153 & UCLA & 18 & 128 & 110 & strong \\
154 & Canterbury Christ Church U. & 1353 & 1240 & 113 & tail \\
155 & Rutgers U.\ New Brunswick & 322 & 208 & 115 & mid \\
156 & Stellenbosch Univ. & 326 & 443 & 117 & mid \\
157 & Case Western Reserve U. & 148 & 265 & 117 & mid \\
158 & Glasgow Caledonian U. & 913 & 795 & 118 & tail \\
159 & Portland State Univ. & 985 & 868 & 118 & tail \\
160 & Sungkyunkwan Univ. & 87 & 205 & 118 & strong \\
161 & Univ.\ of Edinburgh & 29 & 148 & 119 & strong \\
162 & Stockholm University & 204 & 85 & 119 & mid \\
163 & Univ.\ of Sheffield & 111 & 230 & 119 & strong \\
164 & Coventry University & 685 & 805 & 120 & lower \\
165 & Leiden University & 70 & 190 & 120 & strong \\
166 & European U.\ Cyprus & 854 & 975 & 121 & tail \\
167 & Macau U.\ Sci.\ \& Tech. & 260 & 383 & 123 & mid \\
168 & James Cook University & 357 & 480 & 123 & mid \\
169 & Univ.\ of Nizwa & 500 & 623 & 123 & lower \\
170 & Lund University & 96 & 220 & 124 & strong \\
171 & Univ.\ of Granada & 642 & 518 & 125 & lower \\
172 & Yazd University & 1340 & 1465 & 125 & tail \\
173 & Univ.\ of Bath & 263 & 390 & 127 & mid \\
174 & Univ.\ of Pavia & 368 & 240 & 128 & lower \\
175 & Univ.\ of Florida & 135 & 265 & 130 & mid \\
176 & Tech.\ U.\ Denmark & 121 & 253 & 132 & strong \\
177 & Cardiff University & 232 & 100 & 132 & mid \\
178 & Univ.\ of Leeds & 118 & 250 & 132 & strong \\
179 & Aarhus University & 101 & 238 & 137 & strong \\
180 & U.\ Teknologi Brunei & 758 & 895 & 137 & lower \\
181 & Univ.\ of Sharjah & 338 & 475 & 137 & lower \\
182 & Sunway University & 307 & 445 & 138 & mid \\
183 & UC San Diego & 47 & 185 & 138 & strong \\
184 & Yangzhou University & 530 & 670 & 140 & lower \\
185 & Univ.\ of Houston & 459 & 318 & 142 & lower \\
186 & Eindhoven U.\ Tech. & 194 & 340 & 146 & mid \\
187 & Maastricht University & 131 & 278 & 147 & mid \\
188 & NC State University & 303 & 450 & 147 & mid \\
189 & Arizona State Univ. & 213 & 360 & 147 & mid \\
190 & Reichman University & 999 & 850 & 149 & tail \\
191 & Azerbaijan State Oil U. & 1962 & 2115 & 153 & tail \\
192 & Keele University & 563 & 410 & 153 & lower \\
193 & Princess Nourah U. & 504 & 660 & 156 & lower \\
194 & Univ.\ of Salamanca & 916 & 1073 & 157 & lower \\
195 & King Abdulaziz Univ. & 399 & 243 & 157 & lower \\
196 & Swansea University & 321 & 483 & 162 & lower \\
197 & Harbin Inst.\ Tech. & 132 & 305 & 173 & mid \\
198 & Univ.\ of W\"{u}rzburg & 180 & 355 & 175 & mid \\
199 & Macquarie University & 169 & 345 & 176 & mid \\
200 & Peking University & 13 & 190 & 177 & elite \\
201 & Univ.\ of Connecticut & 395 & 215 & 180 & lower \\
202 & TU Dresden & 176 & 358 & 182 & mid \\
203 & Univ.\ of Warwick & 123 & 305 & 182 & strong \\
204 & Ruhr U.\ Bochum & 271 & 455 & 184 & mid \\
205 & Tech.\ U.\ Munich & 27 & 213 & 186 & strong \\
206 & Lebanese American U. & 268 & 455 & 187 & lower \\
207 & Univ.\ of S.\ Denmark & 281 & 470 & 189 & mid \\
208 & Charit\'{e} Berlin & 91 & 280 & 189 & strong \\
209 & Univ.\ of Helsinki & 104 & 295 & 191 & strong \\
210 & Claude Bernard U.\ Lyon 1 & 682 & 490 & 192 & lower \\
211 & Univ.\ of Tabuk & 780 & 588 & 193 & tail \\
212 & Wageningen U.\& Research & 66 & 263 & 197 & strong \\
213 & Hyogo Medical Univ. & 1303 & 1500 & 197 & tail \\
214 & Chiba University & 1055 & 858 & 198 & tail \\
215 & Erasmus U.\ Rotterdam & 107 & 305 & 198 & strong \\
216 & LUT University & 301 & 500 & 199 & mid \\
217 & Univ.\ of Saskatchewan & 397 & 600 & 203 & lower \\
218 & Nanjing U.\ Sci.\ \& Tech. & 622 & 418 & 205 & lower \\
219 & Graz U.\ of Tech. & 633 & 840 & 207 & lower \\
220 & U.\ of Central Florida & 478 & 685 & 207 & lower \\
221 & Univ.\ Sunshine Coast & 578 & 785 & 207 & tail \\
222 & U.\ Tenaga Nasional & 673 & 465 & 208 & lower \\
223 & Univ.\ of Genoa & 410 & 200 & 210 & lower \\
224 & York University & 422 & 635 & 213 & lower \\
225 & Univ.\ of Twente & 191 & 405 & 214 & mid \\
226 & Duke University & 28 & 243 & 215 & strong \\
227 & Univ.\ of Cologne & 165 & 383 & 218 & mid \\
228 & UC Irvine & 97 & 315 & 218 & mid \\
229 & Univ.\ of Bologna & 130 & 350 & 220 & mid \\
230 & KTH & 99 & 320 & 221 & strong \\
231 & Auburn University & 644 & 870 & 226 & lower \\
232 & Yonsei University & 86 & 320 & 234 & strong \\
233 & American U.\ Sharjah & 560 & 795 & 235 & lower \\
234 & London South Bank U. & 713 & 950 & 237 & tail \\
235 & Univ.\ of Valencia & 543 & 303 & 241 & lower \\
236 & Fudan University & 36 & 278 & 242 & strong \\
237 & Auckland U.\ Tech. & 587 & 345 & 242 & lower \\
238 & Univ.\ of T\"{u}bingen & 98 & 345 & 247 & strong \\
239 & Univ.\ of Johannesburg & 388 & 635 & 247 & lower \\
240 & TU Dortmund & 558 & 805 & 247 & lower \\
241 & Chinese U.\ Hong Kong & 44 & 305 & 261 & strong \\
242 & Hebrew U.\ Jerusalem & 261 & 525 & 264 & mid \\
243 & Prince Sultan Univ. & 390 & 650 & 260 & lower \\
244 & Univ.\ of Haifa & 607 & 875 & 268 & lower \\
245 & Ontario Tech Univ. & 947 & 1215 & 268 & tail \\
246 & Izmir Inst.\ Tech. & 1157 & 1425 & 268 & tail \\
247 & Lahore College Women U. & 1417 & 1690 & 273 & tail \\
248 & Semmelweis University & 273 & 555 & 282 & lower \\
249 & U.\ du Qu\'{e}bec & 541 & 828 & 287 & lower \\
250 & Sabanc{\i} University & 393 & 685 & 292 & mid \\
251 & U.\ Sains Malaysia & 412 & 705 & 293 & mid \\
252 & Univ.\ of Windsor & 536 & 830 & 294 & lower \\
253 & Univ.\ of Iceland & 521 & 223 & 299 & lower \\
254 & Yale University & 10 & 310 & 300 & elite \\
255 & Univ.\ of Marburg & 409 & 710 & 301 & mid \\
256 & Universit\'{e} Laval & 413 & 715 & 302 & lower \\
257 & Univ.\ of Duisburg-Essen & 312 & 620 & 308 & mid \\
258 & Univ.\ of Crete & 649 & 340 & 309 & tail \\
259 & Natl.\ Yang Ming Chiao Tung U. & 436 & 748 & 312 & mid \\
260 & Riphah Intl.\ Univ. & 1038 & 1350 & 312 & tail \\
261 & Univ.\ Paris-Saclay & 68 & 383 & 315 & strong \\
262 & Southern Cross Univ. & 499 & 815 & 316 & lower \\
263 & U.\ Brunei Darussalam & 374 & 690 & 316 & mid \\
264 & Univ.\ of Ilorin & 1232 & 1550 & 318 & tail \\
265 & Dhofar University & 677 & 995 & 318 & tail \\
266 & Univ.\ of Konstanz & 283 & 603 & 320 & mid \\
267 & Aalto University & 196 & 518 & 322 & mid \\
268 & Univ.\ of Kansas & 375 & 698 & 323 & mid \\
269 & Univ.\ of Oregon & 458 & 135 & 323 & lower \\
270 & Univ.\ of Sadat City & 1145 & 1475 & 330 & tail \\
271 & SW Jiaotong University & 836 & 505 & 331 & lower \\
272 & Indian Inst.\ Tech.\ Indore & 573 & 905 & 332 & lower \\
273 & Univ.\ of Notre Dame & 195 & 533 & 338 & mid \\
274 & Ohio State University & 108 & 450 & 342 & strong \\
275 & Birla Inst.\ Tech. & 1122 & 1465 & 343 & tail \\
276 & Univ.\ of Guelph & 446 & 795 & 349 & lower \\
277 & Univ.\ of Tokyo & 26 & 378 & 352 & strong \\
278 & Natl.\ Res.\ Nucl.\ U.\ MEPhI & 702 & 1055 & 353 & lower \\
279 & George Washington U. & 226 & 585 & 359 & mid \\
280 & Univ.\ of Reading & 201 & 570 & 369 & mid \\
281 & Masaryk University & 697 & 1068 & 371 & lower \\
282 & Univ.\ of Fribourg & 423 & 800 & 377 & lower \\
283 & Arab Acad.\ Sci.\ Tech. & 1020 & 1403 & 383 & tail \\
284 & Karunya Inst.\ Tech. & 1205 & 1588 & 383 & tail \\
285 & Ajman University & 411 & 795 & 384 & lower \\
286 & Univ.\ of Salford & 851 & 1240 & 389 & tail \\
287 & Wenzhou University & 872 & 480 & 392 & tail \\
288 & Kyoto University & 61 & 455 & 394 & strong \\
289 & Kyushu University & 319 & 720 & 401 & mid \\
290 & Hacettepe University & 877 & 1285 & 408 & tail \\
291 & Kermanshah U.\ Med.\ Sci. & 387 & 800 & 413 & mid \\
292 & Goldsmiths London & 575 & 990 & 415 & lower \\
293 & Saarland University & 522 & 950 & 428 & lower \\
294 & Univ.\ of Nicosia & 514 & 945 & 431 & lower \\
295 & Zhejiang Chinese Med.\ U. & 1451 & 1015 & 436 & tail \\
296 & San Diego State U. & 1179 & 735 & 444 & tail \\
297 & UiT Arctic U.\ Norway & 669 & 1115 & 446 & lower \\
298 & Swinburne U.\ Tech. & 256 & 705 & 449 & mid \\
299 & Bilkent University & 631 & 1083 & 452 & lower \\
300 & Univ.\ of Vienna & 95 & 555 & 460 & strong \\
301 & Abu Dhabi University & 233 & 695 & 462 & mid \\
302 & North South University & 939 & 1405 & 466 & tail \\
303 & Natl.\ Taiwan Normal U. & 570 & 1050 & 480 & lower \\
304 & Worcester Polytechnic Inst. & 721 & 1205 & 484 & tail \\
305 & Victoria U.\ Wellington & 434 & 925 & 491 & mid \\
306 & Zayed University & 497 & 1000 & 503 & lower \\
307 & \'{E}cole Normale Sup.\ Lyon & 315 & 830 & 515 & mid \\
308 & Univ.\ of Malakand & 853 & 1368 & 515 & tail \\
309 & Purdue University & 85 & 605 & 520 & strong \\
310 & Univ.\ of Osaka & 152 & 680 & 528 & strong \\
311 & Federal U.\ Rio de Janeiro & 723 & 1253 & 530 & lower \\
312 & Univ.\ of Luxembourg & 265 & 813 & 548 & lower \\
313 & Edge Hill University & 1092 & 1645 & 553 & tail \\
314 & Prince Mohammad Bin Fahd U. & 352 & 905 & 553 & mid \\
315 & Tohoku University & 105 & 660 & 555 & strong \\
316 & Univ.\ of Zagreb & 1269 & 700 & 569 & tail \\
317 & Gazipur Agri.\ Univ. & 935 & 1513 & 578 & tail \\
318 & Nanjing Tech Univ. & 794 & 213 & 582 & lower \\
319 & Ko\c{c} University & 351 & 953 & 602 & mid \\
320 & La Trobe University & 264 & 880 & 616 & mid \\
321 & Paderborn University & 690 & 1308 & 618 & lower \\
322 & Indraprastha Inst.\ Info.\ Tech. & 1169 & 1800 & 631 & tail \\
323 & Michigan State Univ. & 106 & 758 & 652 & mid \\
324 & Tufts University & 190 & 868 & 678 & mid \\
325 & Univ.\ of Klagenfurt & 637 & 1305 & 668 & lower \\
326 & Chung Yuan Christian U. & 1364 & 670 & 694 & tail \\
327 & Hangzhou Normal Univ. & 1083 & 390 & 693 & tail \\
328 & Ain Shams University & 904 & 1605 & 701 & tail \\
329 & Univ.\ of Colombo & 1008 & 1760 & 752 & tail \\
330 & Indian Inst.\ Science & 251 & 1005 & 754 & mid \\
331 & Univ.\ of G\"{o}ttingen & 122 & 895 & 773 & strong \\
332 & Kazan Federal Univ. & 989 & 1790 & 801 & tail \\
333 & Mahatma Gandhi Univ. & 567 & 1395 & 828 & lower \\
334 & George Mason Univ. & 439 & 1320 & 881 & lower \\
335 & Istanbul Medipol Univ. & 843 & 1733 & 890 & tail \\
336 & Centrale Nantes & 621 & 1563 & 942 & lower \\
337 & Penn State & 110 & 1095 & 985 & strong \\
338 & Univ.\ of Michigan & 23 & 1005 & 982 & strong \\
339 & Univ.\ of Graz & 585 & 1625 & 1040 & lower \\
340 & Univ.\ of Hagen & 891 & 2000 & 1109 & tail \\
341 & ENTPE & 635 & 1790 & 1155 & lower \\
342 & Lomonosov Moscow State U. & 133 & 1400 & 1267 & strong \\
343 & Sciences Po & 612 & 2120 & 1508 & lower \\
344 & MIPT & 370 & 1903 & 1533 & lower \\
345 & Univ.\ of Milan & 323 & 1880 & 1557 & mid \\
346 & Inst.\ Polytechnique de Paris & 69 & 1700 & 1631 & strong \\
347 & Bauman Moscow State Tech.\ U. & 340 & 2080 & 1740 & mid \\
348 & UT Austin & 50 & 1813 & 1763 & strong \\
349 & Friedrich Schiller U.\ Jena & 203 & 2070 & 1867 & mid \\
350 & Univ.\ of Toulouse & 546 & 2425 & 1879 & lower \\
351 & Univ.\ of Cape Town & 166 & 2200 & 2034 & mid \\
\end{longtable}
}

\begin{figure*}[htbp]
  \centering
  \includegraphics[width=\textwidth]{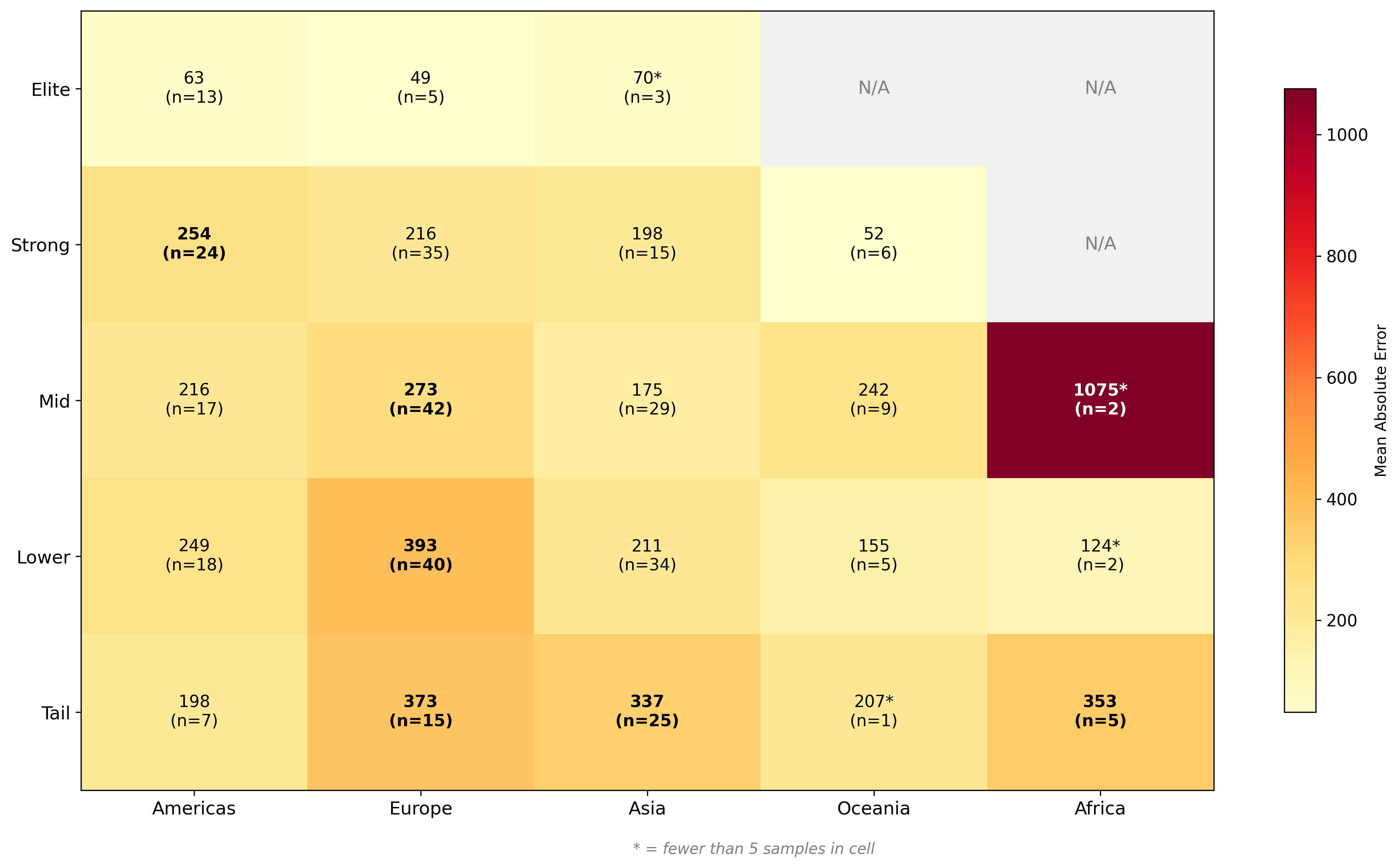}
  \caption{Error heatmap (tier $\times$ region). Cell values show mean AE. Darker cells indicate higher error. Cells with $N < 5$ are marked with an asterisk.}
  \label{fig:heatmap}
\end{figure*}

\end{document}